\documentclass{article}

\pdfoutput=1

\usepackage{arxiv}

\usepackage[utf8]{inputenc} % allow utf-8 input
\usepackage[T1]{fontenc}    % use 8-bit T1 fonts
\usepackage{hyperref}       % hyperlinks
\usepackage{url}            % simple URL typesetting
\usepackage{booktabs}       % professional-quality tables
\usepackage{amsfonts}       % blackboard math symbols
\usepackage{nicefrac}       % compact symbols for 1/2, etc.
\usepackage{microtype}      % microtypography

{}
{}
{}

\usepackage{amssymb,amsmath}
\usepackage{comment}
\usepackage{epsfig} % for postscript graphics files
\usepackage{epstopdf}
\usepackage{color}
\usepackage{losymbol}
\usepackage{float}
\usepackage{bm}
\usepackage[ruled]{algorithm2e}
\usepackage{xcolor}
\usepackage{nomencl}
\usepackage{float}
\usepackage{booktabs}
\usepackage{multicol}
\usepackage{multirow}
\usepackage{subcaption}

\title{MARL-FWC: Optimal Coordination of Freeway Traffic Control Measures}

\author{
  Ahmed Fares\\
  Cyber-Physical Systems Lab\\
  Egypt-Japan University of Science and Technology (E-JUST)\\
  New Borg El-Arab City, Postal Code 21934, Alexandria, Egypt \\
  Faculty of Engineering (Shoubra), Benha University, Cairo 11240, Egypt\\
  \texttt{ahmed.fares@feng.bu.edu.eg} \\
   \And
  Walid Gomaa \\
  Cyber-Physical Systems Lab\\
  Egypt-Japan University of Science and Technology (E-JUST)\\
  New Borg El-Arab City, Postal Code 21934, Alexandria, Egypt\\
  Faculty of Engineering, Alexandria University, Alexandria 21544, Egypt\\
  \texttt{walid.gomaa@ejust.edu.eg} \\
  \AND
  Mohamed A. Khamis \\
  Cyber-Physical Systems Lab\\
  Egypt-Japan University of Science and Technology (E-JUST)\\
  New Borg El-Arab City, Postal Code 21934, Alexandria, Egypt\\
  \texttt{mohamed.khamis@ejust.edu.eg} \\
}

\begin{document}

%A strong manuscript has clear, useful, and exciting message.
%Write with clarity, objectivity, accuracy, and brevity.

%Write direct and short sentences.
%One idea or piece of information per sentence is sufficient.
%Avoid multiple statements in one sentence.

%Present tense for known facts and hypothesis.

%Use active voice to shorten sentences.
%Avoid abbreviations: "it's", "weren't", "hasn't"

%Minimize use of adverbs: "However", "In addition", "Moreover".

% Make the title, abstract, keywords easy for indexing and searching! (informative, attractive, effective)

% A good title should contain the fewest possible words that adequately describe the content of a paper
% Effective titles:
% - Identify the main issue of the paper
% - Begin with the subject of the paper
% - Are accurate, unambiguous, specific, and complete
% - Are as short as possible
% Articles with short, catchy titles are often better cited
% Do not contain rarely-used abbreviations

\maketitle

\begin{abstract}
%\note{abstract have to be only about two paragraphs not many separate paragraphs}
%\note{must increase the portion of taking about framework including our new contribution "in specific" rather than the first two paragraphs of background, i.e., minimize the abstract and focus on new work}

%\boldmath
%The abstract goes here.

%% Text of abstract
% A concise and factual abstract is required. The abstract should state briefly the purpose of the research, the principal results and major conclusions. An abstract is often presented separately from the article, so it must be able to stand alone. For this reason, References should be avoided, but if essential, then cite the author(s) and year(s). Also, non-standard or uncommon abbreviations should be avoided, but if essential they must be defined at their first mention in the abstract itself.

% Make it interesting, and easy to be understood without reading the whole article.
% You must be accurate and specific!
% A clear abstract will strongly influence whether or not your work is further considered.
% Keep it as brief as possible!!!

%Traffic congestion is a challenging problem faced in everyday life. It has multiple effects, such as reduced average speed, accumulated travel time, capacity, and fuel consumption. In addition, it increases the risk factors of accidents and air pollution.
%Hence, drivers could benefit from an intelligent and reliable traffic control system.
The objective of this article is to optimize the overall traffic flow on freeways using multiple ramp metering controls plus its complementary Dynamic Speed Limits (DSLs). An optimal freeway operation can be reached when minimizing the difference between the freeway density and the critical ratio for maximum traffic flow.
In this article, a Multi-Agent Reinforcement Learning for Freeways Control (MARL-FWC) system for ramps metering and DSLs is proposed.
MARL-FWC introduces a new microscopic framework at the network level based on collaborative Markov Decision Process modeling (Markov game) and an associated cooperative Q-learning algorithm. The technique incorporates payoff propagation (Max-Plus algorithm) under the coordination graphs framework, particularly suited for optimal control purposes. MARL-FWC provides three control designs: fully independent, fully distributed, and centralized; suited for different network architectures.
MARL-FWC was extensively tested in order to assess the proposed model of the joint payoff, as well as the global payoff. Experiments are conducted with heavy traffic flow under the renowned VISSIM traffic simulator to evaluate MARL-FWC.
The experimental results show a significant decrease in the total travel time and an increase in the average speed (when compared with the base case) while maintaining an optimal traffic flow.
\end{abstract}

%Immediately after the abstract, provide a maximum of 6 keywords, using American spelling and avoiding general and plural terms and multiple concepts (avoid, for example, 'and', 'of').
%Be sparing with abbreviations: only abbreviations firmly established in the field may be eligible (e.g. DNA)
%These keywords will be used for indexing purposes.

%% main text (IMRAD)
% Introduction, Methods, Results And Discussions
% Journal space is not unlimited
% Make your article as concise as possible

\section{Introduction}
\label{sec:introduction}

%State the objectives of the work and provide an adequate background, avoiding a detailed literature survey or a summary of the results.

% Provide context to convince readers that you clearly know why your work is useful
% Be brief
% Clearly address the following:
% - What is the problem.
% - Are there any existing solutions.
% - Which solution is the best?
% - What is its limitation?
% - What do you hope to achieve?
% Try to be consistent with the nature of the journal.

The  world  is  experiencing  a  period  of  rapid  urbanization,  with  more  than  60  percent  of  the  world population expected to live in cities by 2025 \cite{6582403}.
%This situation creates both unprecedented economic opportunities and challenges.
%According to the assessment of IBM \cite{dirks2010smarter}, cities face four high-impact areas of improvement:
%reducing congestion in transport systems,
%improving public safety and emergency response time,
%improving education and training, and
%enabling access to health care.
The huge  increase in the number and length of traffic jams on freeways has led to the use of several dynamic traffic management measures. The sophistication of traffic network demands, as well as their severity, have also increased recently. Consequently, the need for an optimal and reliable traffic control, for urban freeways networks, has become more and more critical.

Generally, there are three freeway  control strategies: on-ramp control, mainline control, and DSLs control.
%Controlling the number of vehicles entering the freeway from the ramp is called \textit{the rate of ramp metering}, which could be measured by a ramp metering device. %as shown in Fig. \ref{fig:Traffic network with single ramp}:
%O1: main flow origin, O2: ramp flow origin, D1: destination for O1 and O2, S1: road segment before ramp, S2: road segment after ramp, RM: Ramp metering.
Two signals are only generated by a ramp metering device: red and green (no yellow), controlled by a smart or basic controller that regulates the flow of traffic entering the freeway according to the current traffic volume. We aim to control the number of vehicles entering the mainstream freeway from the ramp merging area.  %balance the demand and the capacity of the freeway.
This optimizes the freeway density below the critical one.
%Consequently, this leads to maximum utilization of the freeway without entering in congestion while maintaining the optimal freeway operation.
In order to achieve this flow, \textit{optimal coordination} of the freeway traffic control \textit{measures} over the network level is highly needed.
%\begin{figure}[H]
%\includegraphics[width=\linewidth]{SingleRM.pdf}
%\caption{Traffic network with single ramp and ramp metering as control measure.}
%\label{fig:Traffic network with single ramp}
%\end{figure}

Machine Learning (ML) is one of the fastest growing areas of science.
It has been been used in many applications;
e.g., traffic signal control \cite{Khamis2014AdaptiveMultiObjectiveReinforcementLearning, Khamis2012EnhancedMultiObjectiveTrafficControl,
Khamis2012MultiObjectiveTrafficControl,Khamis2012AdaptiveTrafficControl}, car-following models
\cite{Zaky2015CarFollowingMarkovRegimeClassification}, bioinformatics \cite{Khamis2015MachineLearningInComputationalDocking,Khamis2015ComparativeAssessmentMLScoringFunctionsOnPDBbind2013}, etc.
Reinforcement Learning (RL) is a machine learning technique that addresses the inquiry of how an autonomous agent can learn behavior through trial and error interaction with the dynamic environment to accomplish its goals \cite{kaelbling1996reinforcement}. In order to solve the problem of optimal metering rate on the network level; a \textit{collaborative Multi-Agent System} (MAS) \cite{vlassis2004anytime} based on reinforcement learning is appropriate. A collaborative MAS is defined as the  system  in which the  agents  are designed  to act together  in order to optimize  the sequence of actions. The framework of the Coordination Graphs (CGs) \cite{guestrin2001multiagent} is utilized which is designed based on the dependencies between agents. It decomposes the global payoff function into a sum of local payoffs based on that dependency.

%\indent By using optimal ramp metering control, the freeway flow  can be improved. As a result, the overall total travel time (TTT), fuel consumption, accidents, and air pollution are all improved in quality.
%-------------------------------------------------------------------------------------------------
%\section{Challenges and Motivation}
%\label{sec:Motivation}
%Challenges of the traffic control problem come from the inherent dynamic nature, non-linearity, and uncertainty of the traffic network system \cite{6484604}. Such kinds of problems cannot be locally solved. Hence,  an optimal control sequence over all control decision points is needed. Congestion on a particular section of the freeway is reached when the demand flow exceeds the capacity of this section. It has been broadly agreed that this complex problem cannot be represented exactly using a mathematical model \cite{6484604,4659504}. This is primarily due to the tremendous complexity of the problem and the huge amount of parameters that can contribute to the control actions.

Traffic congestion is a challenging problem faced in everyday life.
It has multiple negative effects on the average speed, overall total travel time, fuel consumption, safety (primary cause of accidents), and environment (air pollution).
Hence, comes the need for an intelligent reliable traffic control system.
%-------------------------------------------------------------------------------------------------
%\section{Research Objectives and Key Contributions}
%\subsection{Main objectives}
%\begin{itemize}
The objective of this article is to optimize the overall traffic congestion in freeways via multiple ramps metering controls plus its complementary multiple DSLs controls.
Maximize the freeway traffic flow by always keeping the freeway density within a small margin of the critical ratio (which is calculated using the calibrating fundamental diagram of the traffic flow).
Preventing the breakdown of a downstream bottleneck through the regulation of traffic stream speed.
Preventing the capacity drop at the merging area via ramp metering control plus its complementary DSLs control,
i.e., complements the function of RM when the flow is above the critical ratio. %(Fig. \ref{fig:pRMDSL1}).
In this article, a Multi-Agent Reinforcement Learning for Freeways Control (MARL-FWC) system for ramp metering and dynamic speed limit is proposed based on a Reinforcement Learning density Control Agent (RLCA).
MARL-FWC introduces a new microscopic framework at the network level based on collaborative Markov Decision Process (Markov game) modeling and an associated cooperative Q-learning algorithm.
The technique incorporates payoff propagation (Max-Plus algorithm) under the coordination graphs framework, particularly suited for optimal control purposes.
A model for the local payoff, the joint payoff, as well as the global payoff is proposed.
MARL-FWC provides three control designs: fully independent, fully distributed, and centralized; suited for different network architectures.

Preliminary results of this work have been published in \cite{6871097,123456}.
In this article, a more detailed description and improvements on MARL-FWC are provided.
An adaptive objective function plus DSLs are presented.
The new approach for ramp metering plus DSLs have shown considerable improvement over the base case.
In addition, a detailed survey of the state-of-the-art work is presented.

% \section{Outline}
% \begin{figure}[!htb]
% \includegraphics[width=\textwidth]{images/ch1/outline1.pdf}
% \caption{Paper outline.}
% \label{fig:paper outline}
% \end{figure}
% Figure \ref{fig:paper outline} demonstrates the paper outline.
The article is organized as follows.
Section \ref{sec:review-background} briefly illustrates the related work.
%, in addition, it gives a brief introduction to the methodology background, particularly single and multi-agent frameworks; Markov Decision Process (MDP) and Markov Game (MG).
Section \ref{sec:ALGORITHMS} describes the RL approach, particularly the Q-learning algorithm.
In addition, it briefly describes the Max-Plus algorithm and associated CGs framework.
Section \ref{sec:proposed-framework} illustrates the design of the RLCA.
Furthermore, it demonstrates the design of three approaches to solve the freeway congestion problem,
particularly the design of the global and joint payoff.
And finally it depicts the MARL-FWC for adaptive ramp metering plus DSLs.
The VISSIM traffic simulator capabilities are also presented.
Section \ref{sec:expres}  represents the results of applying MARL-FWC to different traffic networks and compares its performance to the base case.
Finally, Section \ref{sec:Conclusion} summarizes the conclusions. %and directions for future work
%--------------------------------------------------------------------------------------------------
%\section{Summary}
%Firstly, a brief introduction about the traffic congestion problem is demonstrated.  The sophistication of network demands, as well as their severity, are also illustrated. In addition, the definition of the ramp metering and reinforcement leaning are introduced.
%Then, challenges and motivation are described. Furthermore, the detailed research objectives and key contribution are shown. Finally, paper organization is presented.

\section{Related Work}
\label{sec:review-background}
%--------------------------------------------------------------------------
%\subsection{Background}
%EJUST: IEEE accepts color graphics in the following formats: EPS, PS, TIFF, Word, PowerPoint, Excel, and PDF. The resolution of a RGB color TIFF file should be 400 dpi.
%Provide sufficient detail to allow the work to be reproduced. Methods already published should be indicated by a reference: only relevant modifications should be described.

% Describe how the problem was studied
% - Include detailed information
% - Do not describe previously published procedures.
% - Identify the equipment and describe materials used.

%Block diagram, Fig. \ref{fig:ITSs} demonstrates several potential approaches to mitigate the traffic congestion in both urban intersections and urban freeways. These approaches move from long term solution to short one. The short term solution includes the intelligent transportation systems (ITSs). ITSs consist of three main techniques, but this article focuses only on the advanced traffic management systems.
Ramp metering and speed limits lay under the advanced traffic management systems and are considered one of the most effective traffic control strategies on the freeways recently. Intelligent ramp metering and DSLs mainly depend on the current traffic state. This gives them advance over the fixed time (predefined time signal) and actuated one (sensor located underground to detect the vehicles and change the signal to green).
%\begin{figure}[!htp]
%\includegraphics[width=\textwidth]{images/ARM/ARM1.pdf}
%\caption{Approaches to mitigate the traffic congestion.}
%\label{fig:ITSs}
%\end{figure}
%--------------------------------------------------------------------------
%\subsection{Literature review}
%\label{subsec:lit-rev}
%Challenges of the traffic control problem come from the inherent dynamic nature, non-linearity, and uncertainty of the traffic network system. Such kinds of problems cannot be locally solved. Hence,  an optimal control sequence over all control decision points is needed. Congestion on a particular section of the freeway is reached when the demand flow exceeds the capacity of this section. It has been broadly agreed that this complex problem cannot be represented exactly using a mathematical model \cite{6484604,4659504}. This is primarily due to the tremendous complexity of the problem and the huge amount of parameters that can contribute to the control actions.
Several different techniques have been recently applied including: Intelligent control theory, fuzzy control \cite{6484604,5117656}, neural network control \cite{5678377,5358206}, and even some hybrid of them ~\cite{5720945}.
% Although there have been some progress with such line of research, these techniques need some expertise to extract the parameters and rules and huge amount of training data. Although, some researches move to use RL \cite{veljanovska2012intelligent,davarynejad2011motorway,ji2009optimal}, they only solve the problem locally. Figuring out a congestion only locally, may lead to that the vehicles run faster into another congestion, whereas still the same amount of vehicles have to pass the bottleneck. As a result, the congestion only moves from one road segment to another, while the average travel time of the freeway  remains the same. So there is a need to solve this problem in the \textit{network level} by \textit{coordinated freeway traffic control}.

In this article, a new framework of ramp metering in the network level is proposed based on modeling by \textit{collaborative Markov Decision Process} (Markov Game) \cite{vlassis2004anytime,guestrin2003planning} and an associated \textit{cooperative Q-learning} algorithm, which is based on \textit{a payoff propagation} algorithm \cite{pearl1988probabilistic} under the CGs framework.
MARL-FWC avoids not only the locality of all techniques mentioned above, but also the computational complexity and the risk of being trapped in local optimum of the Model Predictive Control (MPC) \cite{ernst2009reinforcement}. In addition, the solid knowledge of the system considered to extract the rules as in \textit{the fuzzy control system}.
MARL-FWC has been extensively tested in order to assess the proposed model of the joint payoff, as well as the global payoff.
\section{Single and Multi-Agent Algorithms}
\label{sec:ALGORITHMS}

%\section{Summary}
In this section, RL approach is illustrated. This section describes the Q-learning algorithm and shows the difference between the traditional Q-learning and the modified one (temporal difference algorithm).
Then, this section describes briefly the Max-Plus algorithm.
This algorithm is used to determine the option joint action in connected graph, so the coordination graphs framework is introduced.
Finally, freeway traffic flow model is illustrated in details, particularly the fundamental diagram of the traffic flow.

\subsection{Reinforcement learning}
\label{subsec:reinforcement learning}
%RL is a machine learning technique that addresses the question of how an autonomous agent, that perceives its environment through sensors and acts on it through actuators, can learn to choose optimal actions to achieve its goals. This is called \textit{goal directed learning}.
%RL acts without supervision \cite{Mitchell:1997:ML:541177}. The control agent needs to optimize its behavior according to some objective function.
%This introduces the concept of \textit{reward} received from the environment.
%\begin{figure}[H]
%\centering
%  \includegraphics[width=\linewidth]{RL.png}\\
%  \caption{RL Modeling of ramp metering}\label{fig:Rl}
%\end{figure}

%As shown in Fig. \ref{fig:Rl}, the agent interacts with its environment which is described by a set of states $S$. The agent can choose any action from a set of actions $A$ to perform. An agent learns by trial and error interactions with the surrounding environment.
%Every time the agent chooses an action $ a_t$ to be performed in some state $s_t$, the environment responds by a reward or penalty $r_i $ as an evaluation of the quality of this immediate transition.
%These interactions lead to a sequence of state-action pairs $(s_i, a_i)$ and immediate rewards $r_i$ \cite{sutton1998reinforcement}.

The RLCA is interacting with the traffic network. The agent receives the traffic state from the network detectors. This state consists of the number of vehicles $N_t$ and the current density $\rho_t$. Based on this information the RLCA chooses an action $a_t$. As a consequence of $a_t$; the agent receives a reward $r_t$ that keeps the density of the network around a small neighborhood of the critical density. The aim of the agent is to learn a policy which maps an arbitrary state into an optimal action $\pi :S\longrightarrow A$. An optimal action is an action that optimizes the long-term cumulative discounted reward, thus the optimization is done over an infinite horizon. Given a policy $\pi$, the value function corresponding to $\pi$ is stated by Eq. \ref{Equ:reward}:

\begin{equation}
\label{Equ:reward}
V^{\pi}(s_t)=r_t+\gamma^1 r_{t+1}+\gamma^2 r_{t+2}+....=\sum_{i=1}^{\infty}\gamma^i r_{t+i}
\end{equation}

where $\gamma \in (0,1)$ is the \textit{discount factor} and is necessary for the convergence of the previous formula. A policy $\pi^*$ is optimal if its corresponding value function is optimal as in Eq. \ref{Equ:optimal policy}, that is,
for any control policy $\pi$ and any state $s$ (here it is assumed that the objective function is to be maximized):

\begin{equation}
\label{Equ:optimal policy}
V^{{\pi}^{\ast}}(s)\geq  V^{\pi}(s)
\end{equation}

%--------------------------------------------------------------------------------------------------
\subsection{Q-Learning}
\label{subsec:Q-Learning}
%There are many types of reinforcement learning algorithms that use the same principle. The Q-learning algorithm of Watkins \cite{watkins1992q}, that is assumed in current article, is similar to the dynamical programming paradigm.
A Q-learning agent could learn based  on  experienced action sequence actuated in a Markovian environment.
%(see section \ref{sec:Markov Decision Process}).
In fact, Q-learning is a kind of MDP. %(see Algorithm \ref{algo:Q-learning})
In his proof of convergence, Watkins \cite{watkins1992q} assumed a lookup table to represent the value function \cite{watkins1992q}.

%\vspace{1cm}
%\begin{algorithm}[H]
%%\SetLine
%%\KwData{this text}
%%\KwResult{Q-learning }
%Initialize the lookup table $\hat{Q}(s,a)$ (for example, by random small values)\;
%\ForAll{s  $\in$ S , a $\in$ A}{
%	\Repeat{$\hat{Q}$ converges}{
%		Start in an initial state $s$\;
%			\Repeat{$s$ is a terminal state}{
%				Choose action $a$ and execute\;
%				Receive reward $r$ and successor state $s'$\; 	
%				$\hat{Q}(s ,a) := r(s, a) + \gamma \; max_{a'}\hat{Q}(s' , a')$\;
%				$s := s'$\;
%			}
%	}
%}
%\caption{Q-learning}
%\label{algo:Q-learning}
%\end{algorithm}
%\vspace{1cm}

%In Q-learning, the agent first observes the current state $s_t$, and then it chooses an action and executes it. It receives an immediate reward from the environment and observes the next state $s_{t+1}$. Finally, it updates the $Q$-values based on this last experienced interaction with the environment.
This  algorithm is guaranteed to converge to the optimal $Q$-values  with a probability of one under some conditions: the $Q$-function is represented exactly using a lookup table and each state-action pair is visited infinitely often. After convergence to the optimal $Q$-function $Q^*$, the optimal control policy can be extracted as in Eq. \ref{Equ:optimal control policy}:

\begin{equation}
\label{Equ:optimal control policy}
\pi^*(s) = \operatorname*{arg\,max}_{a \in A} Q^*(s,a)
\end{equation}

And the optimal value function (the function that gives a valuation of the states; it can be viewed as a projection of the $Q$-function over the action space) is represented as in Eq. \ref{Equ:Q optimal policy}:
\begin{equation}
\label{Equ:Q optimal policy}
V^*(s) =  V^{{\pi}^{\ast}}(s)=\max_a Q^*(s , a)
\end{equation}
%-----------------------------------------------------------------------

\subsubsection{A modified Q-learning: Temporal difference algorithm}
As long as there is no \textit{model} of the environment (unknown environment), so the infinite sum of the discount reward is no longer considered as a function of state $s$ only as in Dynamic Programming (DP) value iteration \cite{bertsekas1995dynamic}, %(Subsection \ref{sec:Dynamic Programming})
but as a function of the state action pair $(s,a)$. That is why such a $Q$-function \cite{watkins1992q} is used.
When the agent is in state $s$ and the agent performs action $a$ and its \textit{long-term reward} is $Q (s, a)$;
if each state and action pairs are visited infinitely often; then $Q (s, a)$ converges to $Q^{*} (s, a)$ \cite{kaelbling1996reinforcement}. Based on such long-term reward the optimal policy is calculated using Eq. \eqref{Equ:optimal control policy}.
$Q$-learning converges no matter how the actions during the learning are chosen, as long as every action is selected infinitely often (\textit{fair action selection strategy}).
%But of course the speed of the convergence depends on the action that are used during the learning (\textit{action selection strategy}).
A modification of the traditional $Q$-learning rule is in Eq. \eqref{eq:Q-learning}. This new scheme is called a Temporal Difference (TD) learning where $\alpha$ is the learning rate \cite{kaelbling1996reinforcement}.

\begin{align}\label{eq:Q-learning}
&Q^{new}_t(s,a)=Q^{old}_{t-1}(s,a)+\nonumber\\
&\underbrace{\alpha_t(s,a)}_\text{{learning rate}}\left [ \overbrace{\underbrace{R_{t}(s,a)}_\text{reward}+\underbrace{\gamma}_\text {dis. fac.} \:\max_{a^{'}}Q_{t-1}^{old}(s^{'},a^{'})-Q_{t-1}^{old}(s,a)}^\text{TD-Error} \right ]
\end{align}

For alpha value, the easiest approach is to take a fixed value $\alpha$, but even better to use a \textit{variable} $\alpha$.
Eq. \eqref{eq:alpha} is one possibility where $b (s, a)$ is the number of times the action $a$ has been chosen in the state $s$.
\begin{equation}
\label{eq:alpha}
\alpha_t(s,a)=\frac{1}{1+b_t(s,a)}
\end{equation}
%The question now is; what is the best action selection strategy?
There are two extreme possibilities for the best action selection strategy \cite{kaelbling1996reinforcement}.
One extreme is to always choose the action randomly. This what is called \textit{exploration}; just explore what the environment gives in terms of feedback. In the beginning of the learning process this is a good idea. But after some time when the RLCA has already learned, other alternative can be tried which is to select the best action. This is called \textit{exploitation}. In the present case, a combination of exploitation and exploration are used by using $\varepsilon$-greedy action selection strategy, that is with a certain probability $\varepsilon$; the agent always selects a random action.
% RLCA interacts with the traffic network. The agent receives the traffic state $s_t$ from the network detectors. Such state consists of the number of vehicles $N_t$ in the area of  interest associated with RLCA. Accordingly, the RLCA chooses an action $a_t$ (either green or red). As a consequence of $a_t$, the agent receives a reward $r_t$ as an evaluation of the quality of this immediate transition from $s_t$ to $s_{t+1}$.
The state space, the action space and the reward function of the single control agent are described in \cite{6871097} and Subsection \ref{subsec:intelligent agent model for free way ramp}. The joint action space and the (joint and global) payoff function  of collaborative multi-agent are described in detail in \cite{123456} and Subsection \ref{subsec:Cooperative Multi-Agent and  Q-learning}.
%-------------------------------------------------------------------------------
\subsection{Payoff propagation}
\label{subsec:Payoff propagation}
In this subsection, the problem of dynamics in MASs is discussed and the suggestion that the agents should take the behavior of other agents into account.
\subsubsection{Coordination graphs}
\label{subsubsec:coordination Gragh}
CGs \cite{guestrin2001multiagent} can be demonstrated  as an undirected graph $G = (V,\; E)$, where each node $i \in V$ represents an agent, and an edge $e(i,\; j) \in E$ between agents $i,\; j$ means that a dependency exists between them ($j \in \mathrm{\Gamma }(i)$ and $i \in \mathrm{\Gamma }(j)$). Thus, they need to coordinate their actions. CGs allow for the decomposition of a coordination problem into several smaller sub-problems that are easier to solve. CGs have been represented in the context of probabilistic graphs, where the global function, consisting of many variables, is decomposed into local functions. Each of these local functions depends only on a subset of the variables.
The Maximum A Posteriori (MAP) configuration (see Subsection \ref{subsubsec:The Max-plus}) is then estimated by sending locally optimized messages between connected agents (nodes) in the CGs. Even though the message-passing algorithm is developed for estimating the MAP configuration in a probabilistic graph, it is applicable for estimating the optimal joint action of a group of agents in CGs. Because in both situations, the function that is being optimized is decomposed in local terms \cite{kok2006collaborative}.

In collaborative multi-agent systems, the CGs framework \cite{papageorgiou1991alinea} assumes the action of an agent $i$ only depends on a subset of other agents ($j \in \mathrm{\Gamma}(i)$) which may influence its state. The global payoff function $u(\mathbf{a})$ is then broken down into a
linear combination of local payoff functions $f_{i}(\mathbf{a_i})$. The proposed design of the local and global payoff is found in Subsections \ref{subsubsec:Independent Learners} and \ref{subsubsec:Coordinated Reinforcement Learning With Max-Plus} respectively. The global $Q$-function $Q(\mathbf{s},\; \mathbf{a})$ is then decomposed into a sum of local functions given by:
\begin{equation}
\label{eq:global Q}
Q(\mathbf{s},\;\mathbf{a})=\sum_{i=1}^n Q_i ({s_i},\;\mathbf{a_i})
\end{equation}
Where: $\mathbf{a}= (a_1 ,...,a_n )$ is the joint action resulting from: each agent $i$ selects an action $a_i$ from its action set $A_i$, $\mathbf{s}= (s_1 ,...,s_n ) \in S_1 \times \cdots \times S_n$ is the joint state. The joint $Q$-function $Q_i ({s_i},\mathbf{a_i})$ depends only on $\mathbf{a_i}\subseteq  \mathbf{a}$, where $\mathbf{a_i} \in A_i \times \prod_{j \in \mathrm{\Gamma(i)}} A_j$, $\mathrm{\Gamma (i)}$ is the set of neighborhoods of agent $i$.
The decomposition in Eq. \eqref{eq:global Q} can be demonstrated in Fig. \ref{fig:CG involving 4 agents with only pair-wise dependencies} (an example of a CGs of four agents where the dependence among agents is decomposed into pair-wise functions \cite{yedidia2003understanding}); where $f_{ij}$ is the joint payoff between agent $i$ and agent $j$.
%------------------------------------------------------------------------------
\subsubsection{Max-Plus algorithm}
\label{subsubsec:The Max-plus}
%\ref{algo:max-plus}
The Max-Plus algorithm approximates the optimal joint action by iteratively sending locally optimized messages between connected agents (\textit{neighbors}) in the coordination graph.  It is similar to the belief propagation or the max-product algorithm, which estimates the MAP configuration in Belief Networks \cite{kschischang2001factor}. Max-Plus was originally developed for computing MAP solutions in Bayesian networks \cite{pearl1988probabilistic}. The MAP for each agent is estimated using only the incoming messages from other agents. It has been shown that the message updates converge to a fixed point after a finite number of iterations for  cycle free graphs. The message mainly depends on the decomposition given by Eq. \eqref{eq:global Q} and can be defined as follows:
\begin{equation}
\mu_{ij}(a_j)=\max_{a_i} \left \{ f_i\left ( a_i \right )+f_{ij}(a_i,a_j)+\sum _{k\in \mathrm {\Gamma (i)}\backslash j}\mu_{ki}(a_i)  \right \}
\end{equation}

\begin{figure}[H]
        \centering
        \begin{subfigure}[b]{0.8\linewidth}
                \includegraphics[width=\linewidth]{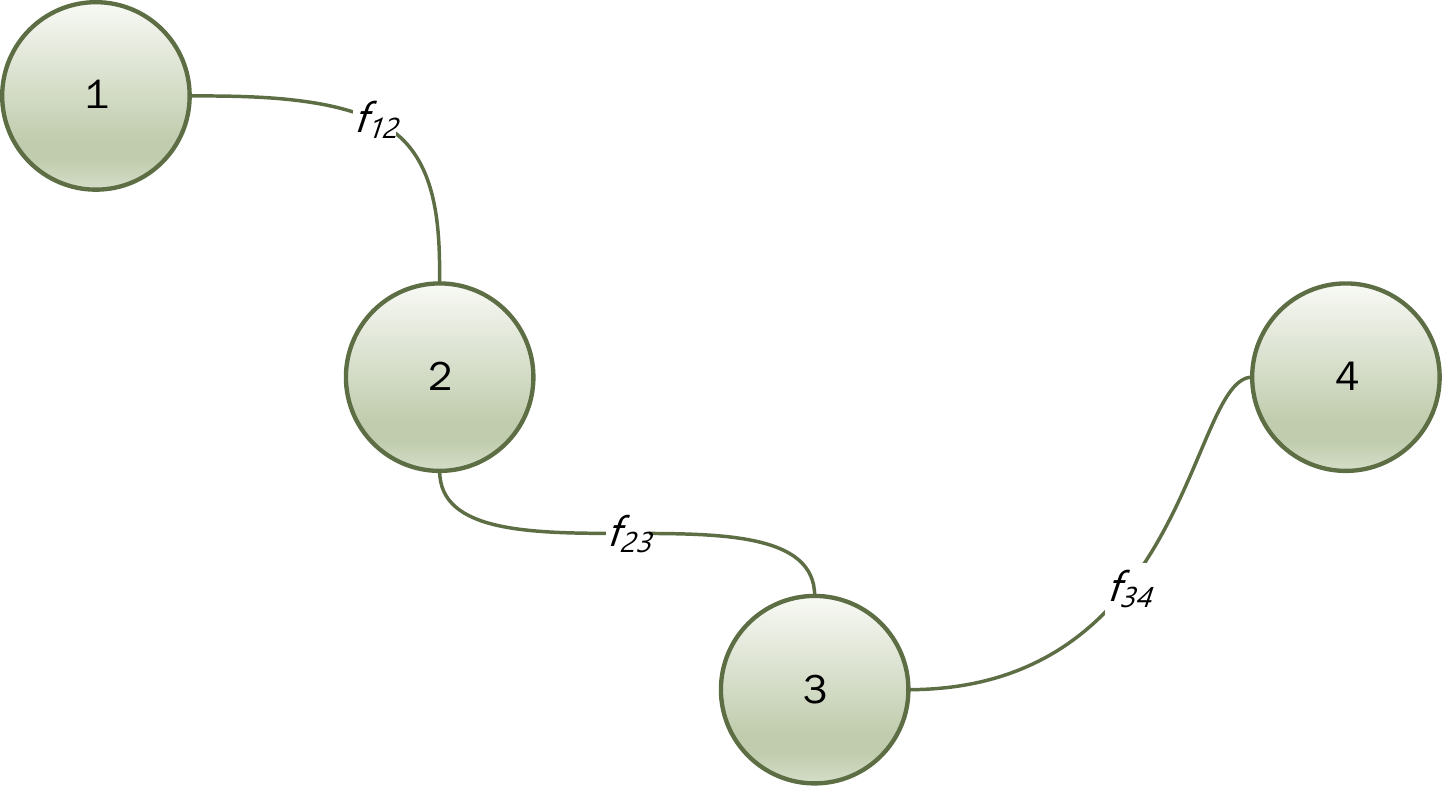}
                \caption{CG involving 4 agents with only pair-wise dependencies.}
                \label{fig:CG involving 4 agents with only pair-wise dependencies}
        \end{subfigure}%
        ~ %add desired spacing between images, e. g. ~, \quad, \qquad etc.
          %(or a blank line to force the subfigure onto a new line)

        \begin{subfigure}[b]{0.8\linewidth}
                \includegraphics[width=\linewidth]{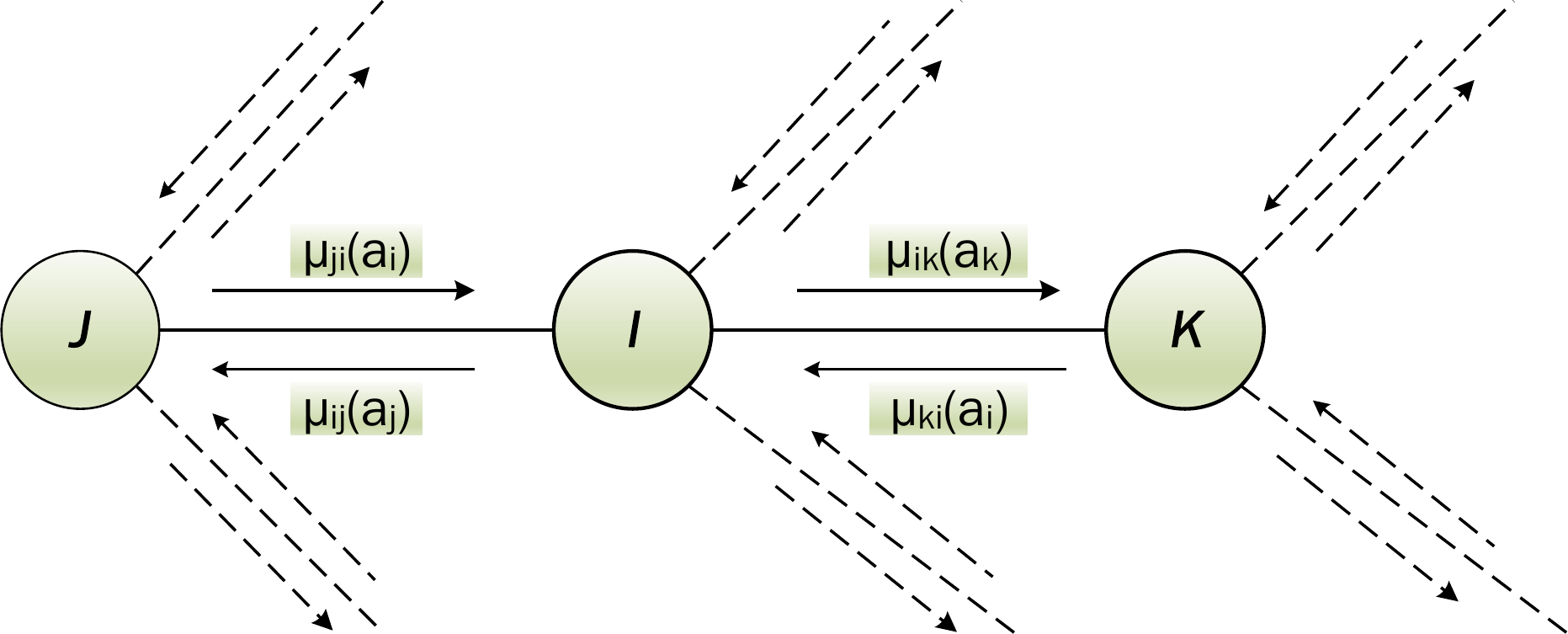}
                \caption{Representation of the different $\mu_{ij}$ sent between neighboring agents.}
                \label{fig:max_plus}
        \end{subfigure}
\caption{Payoff propagation: Coordination graphs and Max plus.}
\label{fig:Payoff propagation}
\end{figure}

Where: $k\in \mathrm {\Gamma (i)}\backslash j$ is the subset of all neighbors  connected to $i$ except $j$, and $\mu_{ij}(a_j)$ is the message from an agent $i$ to agent $j$ after agent $j$ preforms the action $a_j$ as an evaluation of the influence of this action on agent $i$ state. The message approximates the maximum reward agent $i$ can achieve for a committed  action of agent $j$. This message is the max-sum of the local payoff $f_i\left ( a_i \right )$, the joint payoff $f_{ij}(a_i,a_j)$ and all incoming messages agent $i$ received except that from agent $j$. The design of the joint payoff is found in detail in Subsection \ref{subsubsec:Coordinated Reinforcement Learning With Max-Plus}. Figure \ref{fig:max_plus} shows a CG with three agents and the corresponding messages.

If the graph has no cycle, Max-Plus always converges after a finite number of steps to a fixed point in which the messages are changed below a certain threshold \cite{wainwright2004tree}. At each time step, the value for an action $a_i$ of an agent $i$ can be determined as follows:
\begin{equation}
g_i (a_i ) = f_i (a_i )+ \sum _{j\in\Gamma(i)} \mu_{ji} (a_i)
\end{equation}
When the convergence is held, optimal joint action $\mathbf{a}$ has the element $a^\ast_i$ which can be computed as follows:
\begin{equation}
a^\ast_i = \operatorname*{arg\,max}_{a_i} g_i (a_i)
\label{eq:a}
\end{equation}

%\begin{algorithm}[H]
%%\SetLine
%initialization $\forall(i, j)\in E :\;$ $\mu_{ij}$ $\;=\; $ $\mu_{ji} = 0\;$, $\;\forall_{i} \in  V \; : \; g_{i} \; = \;0\;,\;$and$\; m = - \infty$\\
%\While{fixed point $=$ false and deadline to send action has not yet arrived }{
%	$//$ run one iteration\;
%	$fixed\_ \;point \;=\; true$\;
%		\For{every agent $\textbf{\textit{i}}$ }{
%			\ForAll{ neighbors $j \;=\;\mathrm {\Gamma(i)}\;$ }{
%			send $j$ msg $\mu_{ij}(a_j)=\max_{a_i} \left \{ f_i\left ( a_i \right )+f_{ij}(a_i,a_j)+\sum _{k\in \mathrm {\Gamma (i)}\backslash j}\mu_{ki}(a_i)  \right \}$\;
%				\If{$\mu_{ij}\;$differs from previous message by a small threshold}{
%				$fixed\_ \;point \;=\; false$\;
%				}
%			}
%			determine $\;g_{i}(a_i) = f_i(a_i)+\sum _{j\in \mathrm {\Gamma (i)}\backslash j}\mu_{ji}(a_i)\;$and$\;{a'}=\operatorname*{arg\,max}_{a_i} g_{i}(a_i)$\;
%		}
%	
%	\eIf{use anytime extension}{
%	\If{$u(({a'}_{i}))> m$}{
%	$(a^*_i)=(a'_{i})\;$ and $\;m=u((a'_i))$\;
%	}
%	}
%	{$(a^*_i)=(a'_{i})\;$\;}
%}
%\textbf{Return} ($a^*_{i}$)
%\caption{Centralized Max-Plus algorithm for $CG = (V, E)$}
%\label{algo:max-plus}
%\end{algorithm}
%\vspace{1cm}
%
In the Max-Plus algorithm; at each iteration, an agent $i$ sends a message $\mu_{ij}(a_j)$ to each neighbor $j\in \mathrm {\Gamma (i)}$ given the joint payoff $f_{ij}$ and current incoming messages it has received and also computes its current optimal action given all the messages it has received so far that is why it is called \textit{anytime algorithm} \cite{kok2006collaborative}.
The process continues until all messages converge, or a deadline signal is received which means the agents must report their individual actions. In the anytime algorithm, the joint action Eq. \ref{eq:a} is only updated when the corresponding global payoff improves. %The Max-Plus algorithm is extended by computing the global payoff and updating the joint action only when it improves upon the best value found so far. This ensures that every new joint action produces higher payoff than the previous one.

\section{MARL-FWC Architecture}
\label{sec:proposed-framework}
\begin{figure}[H]
\center
\includegraphics[width=\linewidth]{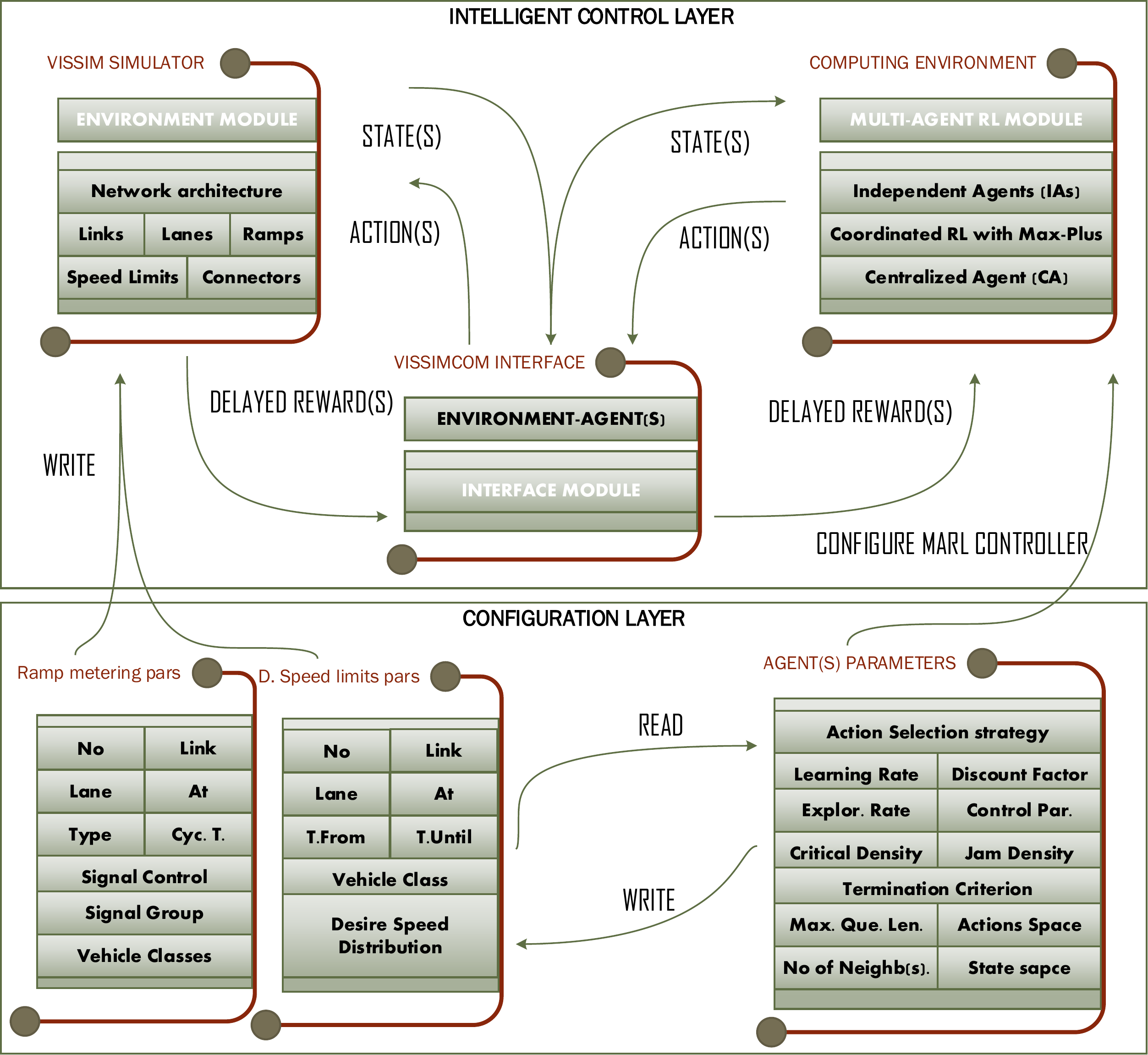}
\caption{MARL-FWC architecture.}
\label{fig:the proposed framework}
\end{figure}

Figure \ref{fig:the proposed framework} demonstrates MARL-FWC architecture.
MARL-FWC consists of two layers; the intelligent control layer and the configuration layer.
The configuration layer consists of three blocks:
\begin{itemize}
\item Ramp metering parameters: configure the ramp metering.
\item DSLs parameters: configure the DSLs.
\item Agents' parameters: configure the control agent.
\end{itemize}
The main task of the configuration layer is to configure the control layer. The control layer consists of three interacting modules (blocks):
\begin{itemize}
\item Environment module (VISSIM simulator): models the traffic network.
\item Multi-agent reinforcement-learning module (Computing environment): implements different controls strategies.
\item Interface module (VISSIM interface): facilitates the interaction between the Multi-agent reinforcement-learning module and the environment module.
\end{itemize}

%------------------------------------------------------------------------
In the following subsections, the main components of the MARL-FWC architecture are described in details,
e.g., RLCA (used in the IAs framework).
\subsection{MARL-FWC with single ramp and RLCA as control measure}
\label{subsec:intelligent agent model for free way ramp}

In the following subsections, all the components of the Markov decision process that corresponds to the on-ramp metering RLCA are described.% see Fig. \ref{fig:Traffic network with single ramp}.
%\begin{figure}
%\includegraphics[width=\textwidth]{images/proposed/SingleRM.pdf}
%\caption{Traffic network with single ramp and RLCA as control measure.}
%\label{fig:Traffic network with single ramp}
%\end{figure}

\subsubsection{State space}
\label{subsubsec:States}
%The state of the freeway network mainly depends on traffic measurements and demands at a given time.
The state space is three-dimensional where each state vector $s_{t+1}$ at time $t+1$ consists of the following components:
%\begin{enumerate}
the number of vehicles in the freeway $N(t+1)$ (see \cite{6871097}), %Eq. \eqref{Equ:NumberOfCar}
the number of vehicles that entered the freeway from ramp $\Delta N(t+1)$, and
the ramp traffic signal at the previous time step $Ts(t)$.
%\end{enumerate}

\subsubsection{Action space}
\label{subsubsec:Action}
%In order to \textit{optimize the density of the freeway}, the RLCA agent should select the optimal actions sequence based on the $Q$-learning algorithm. 
By metering on the on-ramp, the red and green phases of the ramp metering change in order to control the flow entering the freeway from the merging point. So, the action space is modeled as consisting only of two actions: red and green. The optimal action is then chosen based on Eq. \eqref{Equ:optimal control policy}.

\subsubsection{Reward function}
\label{subsubsec:Reward}
%The reward function may either reward or punish the agent's last action.
Since the RLCA's goal is to keep the freeway density $\rho$ around a small neighborhood of the critical density $\rho_{cr}$,
the reward function is designed so as to depend on the current freeway density $\rho$ and how much it deviates from the critical density $\rho_{cr}$.
Hence, the reward of taking an action $a$ in state $s$ is designed as following:

\begin{equation}
\label{Equ:Reward}
Minimize \;\;O=|\rho-\rho_{cr}|: r(s , a)=-|\rho-\rho_{cr}| %\nonumber
\end{equation}

As long as the $Q$-value Eq. \eqref{eq:Q-learning} (as a function of current state and action) is maximized, the difference between the freeway density $\rho$ and the critical density $\rho_{cr}$ is minimized.
Therefore, the best utilization of the freeway is achieved without causing congestion.
%--------------------------------------------------------------------------

\subsection{MARL-FWC based on multi-agent and cooperative Q-learning}
\label{subsec:Cooperative Multi-Agent and  Q-learning}
%In this section, the design of different approaches that control the amount of vehicles entering the mainstream freeway from \textit{ramps} (Fig. \ref{fig:Traffic network with three ramps}) to keep the freeway density with a small margin from the critical ratio is studied. Consequently, this leads to maximum utilization of the freeway without entering in congestion while maintaining the optimal freeway operation.
%\begin{figure}[!htb]
%\includegraphics[width=\textwidth]{images/proposed/MultiRamp.pdf}
%\caption{Traffic network with three ramps.}
%\label{fig:Traffic network with three ramps}
%\end{figure}
\subsubsection{Independent agents}
\label{subsubsec:Independent Learners}
The ``independent learning'' is the first proposed design to solve the freeway congestion problem (Fig. \ref{fig:pIAs}).
%In this design, the agents have a greedy behavior, where each control agent chooses an action (either red or green) that maximizes its local reward.
The $Q$-function is updated as follows:
\begin{align}\label{eq:Independent Learners}
&Q_i^{new}(s_i,a_i)=Q_i^{old}(s_i,a_i)+\nonumber\\
&\alpha \left[R(s_i,a_i)+\gamma \;\max_{a'_{i}}Q_i^{old}(s'_{i},a'_{i})-Q_i^{old}(s_i,a_i)\right ]
\end{align}
Where: $R(s_i,a_i)=-{|\rho(s_i,a_i)-\rho_{cr}|}$ from Eq. \eqref{Equ:Reward}, with such conception of the objective function, the agents try to keep the freeway density within a small margin of the critical ratio; that ensures the maximum utilization of the freeway without entering in congestion. In this $Q$-function design, the agents are partially observing the environment. An agent observes its state which is defined as the number of vehicles in the areas of interest associated with that agent and chooses the local action either red or green, independent of all other agent actions. This cheap design is recommended when the distances between ramps are too long.

\subsubsection{Coordinated reinforcement learning with Max-plus}
\label{subsubsec:Coordinated Reinforcement Learning With Max-Plus}
The second design is considered as a completely distributed design (Fig. \ref{fig:pM+}). It considers a MARL based design. This design works on the network level and is based on modeling by \textit{collaborative Markov Decision Process (Markov Game)} and an associated \textit{cooperative Q-learning} algorithm. This design incorporates \textit{a payoff propagation} (Max-plus algorithm) under the \textit{coordination graphs} framework, particularly suited for optimal control purposes.
In this design,  the control agent coordinates its actions with its neighboring agents (message passing algorithm). The agent updates the cooperative $Q$-function globally as follows:
\begin{align}\label{eq:Coordinated Reinforcement Learning With Max-Plus}
&Q_i^{new}(s_i,\mathbf{a_i})=Q_i^{old}(s_i,\mathbf{a_i})+\nonumber\\
&\alpha \left[R(\textbf{s},\textbf{a})+\gamma \;\max_{\mathbf{a'}}Q^{old}(\mathbf{s'},\mathbf{a'})-Q^{old}(\mathbf{s},\mathbf{a})\right ]
\end{align}
Where:
\begin{itemize}
\item $Q(\mathbf{s},\mathbf{a})$ can be computed using Eq. \eqref{eq:global Q}.
\item $\alpha$ can be computed using Eq. \eqref{eq:alpha}.
\item R(\textbf{s},\textbf{a}) can be figured out by the proposed conception of the global reward function using the \textit{harmonic mean}. With such design, it is guaranteed the balance between the control agents payoffs, and hence optimal response to the freeway dynamics, as follows:
\end{itemize}
\begin{equation}
\label{eq:global reward}
R(\textbf{s},\textbf{a})=\frac{1}{\frac{1}{n}\sum_{i=1}^n\frac{1}{R(s_i,\mathbf{a_i})}}
\end{equation}

The maximum control action $\mathbf{a'}$ in state $\mathbf{s'}$ and its associated estimation of optimal future value $\max_{\mathbf{a'}}Q(\mathbf{s'},\mathbf{a'})$ can be computed using the Max-Plus algorithm.
The function $f_{ij}(a_i,a_j)$ can be computed using the proposed design of the joint payoff between two neighboring control agents (with such design, it is guaranteed the balance between connected control agents), as follows:
\begin{equation}
f_{ij}(a_i,a_j)=\frac{2*f_i(a_i)*f_j(a_j)}{f_i(a_i)+f_j(a_j)}
\end{equation}
This moderate design is recommended when the distances between ramps are short and the traffic network has many ramps.
%--------------------------------------------------------------------------------------
\subsubsection{Centralized agent}
\label{subsubsec:Centerlized Agent}
\begin{figure}[H]
        \centering
        \begin{subfigure}[b]{\linewidth}
                \centering
                \includegraphics[scale=0.7]{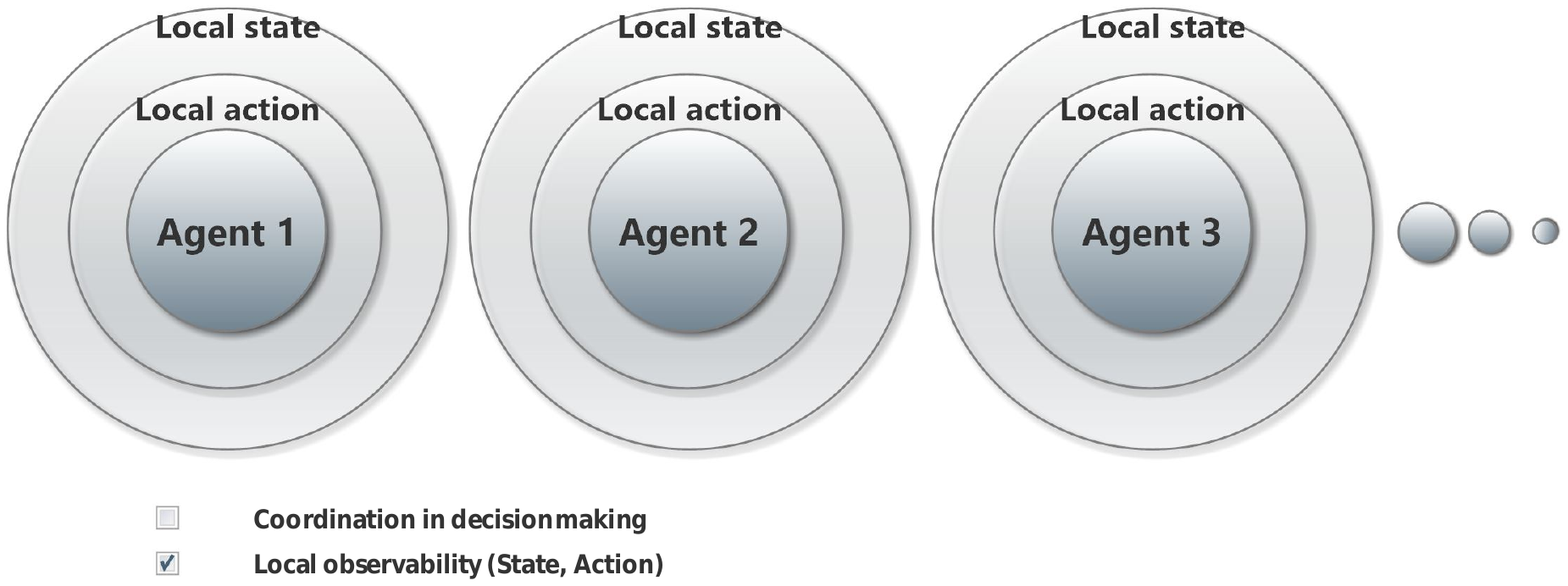}
                \caption{Independent Agents (IAs).}
                \label{fig:pIAs}
        \end{subfigure}%
        ~ %add desired spacing between images, e. g. ~, \quad, \qquad etc.
          %(or a blank line to force the subfigure onto a new line)

        \begin{subfigure}[b]{\linewidth}
                \centering
                \includegraphics[scale=0.7]{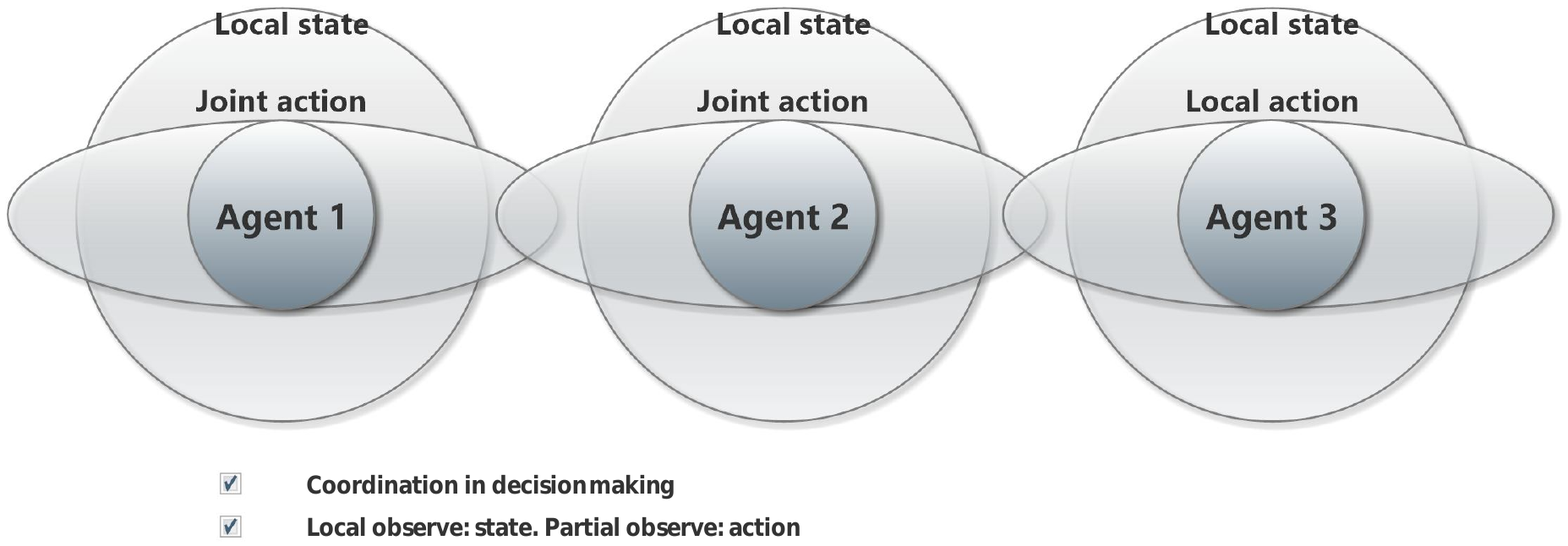}
                \caption{Coordinated reinforcement learning with Max-plus (M+).}
                \label{fig:pM+}
        \end{subfigure}

        \begin{subfigure}[b]{\linewidth}
                \centering
                \includegraphics[scale=0.7]{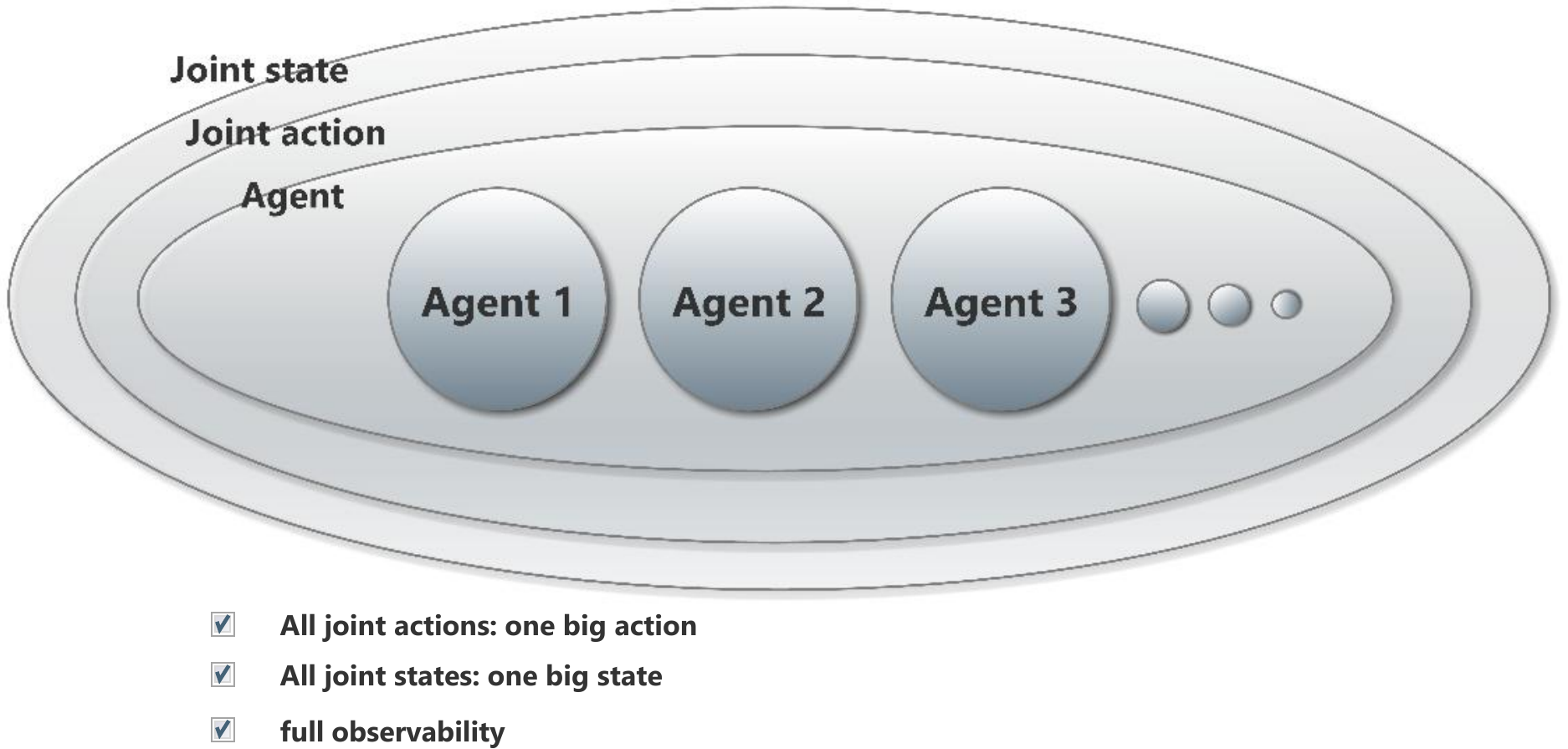}
                \caption{Centralized Agent (CA).}
                \label{fig:pCA}
        \end{subfigure}
\caption{MARL-FWC based on multi-agent and cooperative Q-learning.}
\label{fig:MARL-FWC based on multi-agent and cooperative Q-learning}
\end{figure}

The third extreme design considers a centralized controller (Fig. \ref{fig:pCA}) using a collaborative MAS coordinated actions are orchestrated as a single action. The cooperative $Q$-function for the joint actions are updated using a single $Q$-function, as follows:
\begin{align}
&Q^{new}(\mathbf{s},\mathbf{a})=Q^{old}(\mathbf{s},\mathbf{a})+\nonumber\\
&\alpha \left[R(\textbf{s},\textbf{a})+\gamma \;\max_{\mathbf{a'}}Q^{old}(\mathbf{s'},\mathbf{a'})-Q^{old}(\mathbf{s},\mathbf{a})\right ]
\end{align}
Where: R(\textbf{s},\textbf{a}) can be figured out by Eq. \eqref {eq:global reward}. Nevertheless, this design leads to an optimal solution \cite{watkins1992q}, it is not scalable as it suffers from the curse of dimensionality. This costly design is recommended when the distances between ramps are short and the traffic network has  fewer number of ramps.
%---------------------------------------------------------------------------------------

\subsection{MARL-FWC for adaptive ramp metering plus DSLs}
\label{subsec:Proposed Adaptive Ramp Metering plus Dynamic Speed Limits}
%\begin{figure}[H]
%\includegraphics[width=\linewidth]{RMDSL1.pdf}
%\caption{A combination of DSLs and ramp metering as control measures.}
%\label{fig:pRMDSL}
%\end{figure}
This section illustrates the new conception of ramp metering objective function. This conception takes into account the ramp queue length. This section also demonstrates how DSLs can complement the ramp metering in order to mitigate the traffic congestion when the network is dense. %(Fig. \ref{fig:pRMDSL}).
\subsubsection{MARL-FWC another conception of ramp metering objective function}
\label{subsubsec:New  proposed conception of ramp metering objective function}
By considering the weighted sum of both freeway density and ramp, it is possible to minimize both the normalized difference from the critical density in the freeway $O_h$ and the normalized queue length in the ramp $O_r$.
The ramp metering objective function $O$ is demonstrated as follows:\\
Minimize O:
\begin{align}
\label{eq:new FW objective function}
&O &=&w_h O_h + w_r O_r:&&\text{Collective Objective Function}\\
&O_h &=&2\frac{\left | \rho - \rho_c \right |}{\rho_{max}}:&&\text{Freeway Objective Function}\nonumber \\
& & & &&\text{(normalized difference from critical density)}\nonumber \\
&O_r &=&\frac{q}{q_{max}}:&&\text{Ramp Objective Function}\nonumber \\
& & & &&\text{(normalized queue length)}\nonumber\\
&w_h &=&\frac{\rho \alpha}{\rho_{max}}:&&\text{Adaptive Freeway Weight}\nonumber \\
&w_r &=&\frac{q (1-\alpha)}{q_{max}}:&&\text{Adaptive Ramp Weight}\nonumber \\
&\alpha&\in&[0,1]:&&\text{Control Parameter}\nonumber
\end{align}

Where: $w_h$ is the adaptive freeway weight. It is adaptive because it depends on the current freeway state; as far as the density increases, the $w_h$ also increases. $w_r$ is the adaptive ramp weight. It is adaptive because it depends on the current ramp state; as far as the queue length increases, the $w_r$ also increases. Finally, $\alpha$ is the control parameter. It is used to determine the importance of each weight (biasing  factor). With such conception of ramp metering objective function, there is a balance between mitigating the traffic congestion at the freeway (capacity drop) and moving the problem to another road segment.
%-----------------------------------------------------------------------
\subsubsection{DSLs objective function}
\label{subsubsec:Dynamic speed limits Objective function}
The DSLs analytical proof
%Before moving to how DSLs is designed, there is a need first to
proves that changing the velocity affects the freeway density.
This section illustrates the analytical proof.
This proof is used to demonstrate how the change in velocity can affect both freeway density and flow. It shows that the slope between density and flow is the velocity, see Fig. \ref{fig:Density, flow and velocity relationship}.
\begin{figure}[H]
\centering
\includegraphics[scale=0.75, clip,keepaspectratio]{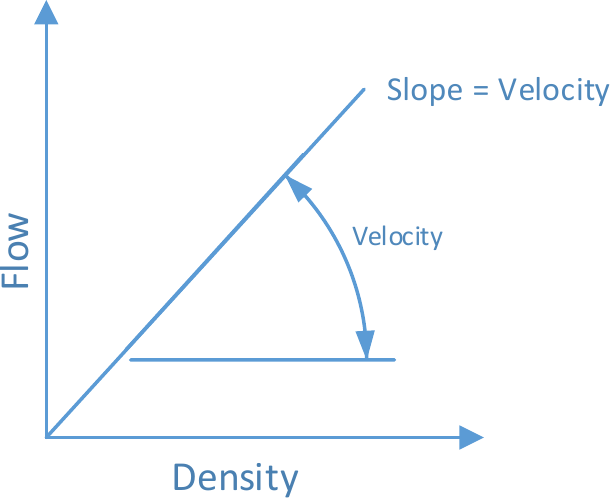}
\caption{Density, flow and velocity relationship.}
\label{fig:Density, flow and velocity relationship}
\end{figure}
\begin{align}
&q&=&\frac{N}{t}&&\text{Freeway Flow (veh/h)}\nonumber \\
&\rho&=&\frac{N}{A}=\frac{N}{(d*width(const.))}&&\text{Freeway Density (veh/km/lane)}\nonumber \\
&v&=&\frac{d}{t}=\frac{{1}/{t}}{{1}/{d}}&&\text{Freeway Velocity (km/h)}\nonumber \\
& & &~\text{Multiply both sides by}~N\nonumber \nonumber \\
&v&=&\frac{{N}/{t}}{{N}/{d}}\nonumber \\
&v&\propto&\frac{q}{\rho}
\end{align}
Where: $N$ is the number of vehicles (veh) (see \cite{6871097}) %Eq. \eqref{Equ:NumberOfCar}
and $t$ is the time (h).
%------------------------------------------------------------------------

DSLs objective function: taking into account the queue length in designing the freeway objective function Eq. \eqref{eq:new FW objective function} raises a question; how to control the amount of vehicles that enter the merging area from the main stream freeway. Hence, the role of the DSLs can be considered as the main stream metering to complement the ramp metering function. When the ramp metering cannot mitigate the capacity drop at the merging area, the DSLs can complement by limiting the number of vehicles entering the merging area. That been said, the DSLs objective function is as follows:
\begin{align}
&O_{hDSL}&=&\quad 2\frac{\left | \rho - \rho_c \right |}{\rho_{max}},\text{Freeway DSLs Objective Function}\nonumber\\
& & &\text{(normalized difference from critical density)}
\end{align}
%------------------------------------------------------------------------

\subsection{The VISSIM simulation environment}
\label{subsec:simulation vissim}
%There exists several types of traffic simulators \cite{maciejewski2010comparison,6582403}.
%VISSIM is considered one of the best technologies for advanced traffic simulation and very well suited for MARL-FWC.
%That is because of its flexibility through \textit{VISSIM COM interface} and \textit{VISSIM DLL API} which provide freedom to traffic researchers and engineers.
In VISSIM there are many functions and parameters which control the VISSIM itself and associated study experiments.  These parameters can be assigned through VISSIM GUI and remain fixed during the running of the experiment or manipulated through programming via VISSIM COM interface which gives the ability to change these parameters during the running time.
For example, the traffic signal during the running time of the experiment can be changed in order to respond to the dynamics of the freeway. VISSIM COM interface can be programmed via any type of programming languages with the ability to handle COM object.
Accordingly, the next section depicts the advanced VISSIM simulator possibilities, particularly ramp control development and DSLs. Some examples of dense networks with different number of ramps are given, however, MARL-FWC can handle any type of networks with any number of ramps.
%----------------------------------------------------------------------------

%\section{Summary}
%Firstly, MARL-FWC architecture is demonstrated. After that, a multi-agent reinforcement-learning control system for ramps metering and DSLs is proposed. Our system introduces a new microscopic framework at the network level based on collaborative Markov Decision Process modeling (Markov game) and an associated cooperative Q-learning algorithm. The technique incorporates payoff propagation (Max-Plus algorithm) under the coordination graphs framework, particularly suited for optimal control purposes. MARL-FWC provides three control designs: fully independent, fully distributed, and centralized; suited for different network architectures. A model for the local payoff, the joint payoff, as well as the global payoff are also proposed.

%Then, this section also shows that when the ramp metering could not mitigate the capacity drop at the merging area, DSLs could complement the ramp metering by adjusting the flow rate on the main street via changing the speed limits' values.

%Finally, this section shows the new conception of the adaptive ramp metering when taking the queue length into account and DSLs. Furthermore, it illustrates the VISSIM simulator capabilities.

\section{Performance Evaluation}
\label{sec:expres}

\subsection{Multi-agent reinforcement learning}
\label{subsec:Experimental-results-Multi-agent-reinforcement-learning}
\begin{figure}[H]
\centering
\includegraphics[width=\linewidth]{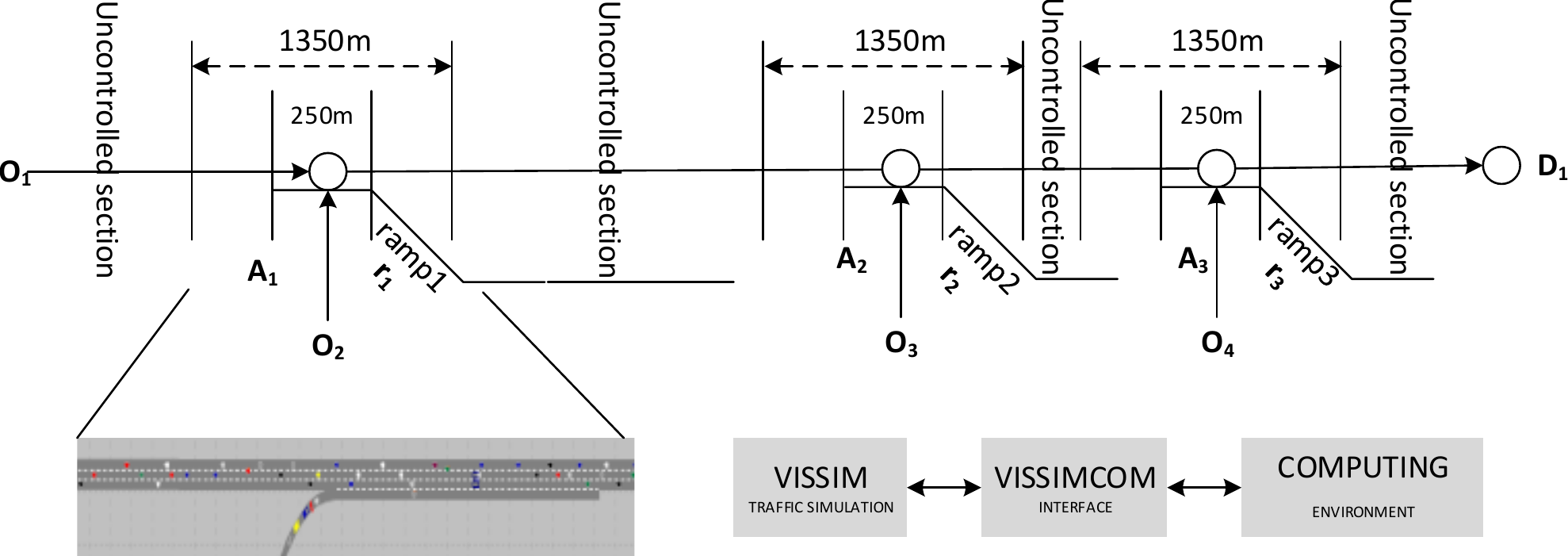}
\caption{The studied network.}
\label{fig:studied network}
\end{figure}

The studied network in Fig. \ref{fig:studied network} consists of a mainstream freeway with three lanes and three-metered on-ramp with one lane each. The network consists of four sources of inflow: $O_1$ for the mainstream flow, $O_2,\;O_3,$ and $O_4$ for the three on-ramp flow. It has only one discharge point $D_1$ with unrestricted outflow.
The mainstream freeway consists of three areas of interest: $A_1$, $A_2,$ and $A_3$, plus three control agents (RLCA) which are located at the entrance points of each ramp. Each area of interest is associated with one control agent. Each area is about $1350\;m$ as follows: $580\;m$ before the ramp, $250\;m$ as a merging area and $520\;m$ after the ramp. There are four uncontrolled sections; three of them before $A_1$, $A_2$ (which is about $1500\;m$) and $A_3$ (which is about $100\;m$) and the fourth is after $A_3$.

Table \ref{tab:demand scenario} represents the demand for both the mainstream freeway and ramps. 
% This demand scenario gives the opportunity to study the effect of the on-ramp smart control. 
That is because with such scenario the density $\rho$ can exceed the critical density $\rho_{cr}$ which is proven during the learning process.
Random choice of action can lead to a scenario with density $\rho$ equals $72$ (Veh/km) which is higher than the critical one $\rho_{cr}$ which is equal to $62$ (Veh/km). The studied network is considered a dense network where the smart control system is highly needed.

%\begin{figure}[!htb]
%\centering
%\includegraphics[scale=0.5]{demand.pdf}
%\caption{The demand scenario.}
%\label{fig:demand scenario}
%\end{figure}

\begin{table}[H]
\centering
\caption{The demand scenario.}
\label{tab:demand scenario}
\begin{tabular}{l|c|c|c|c|c|c|c}
\hline
Simulation Time (sec)       & 300  & 600  & 900  & 1200 & 1500 & 1800 & 2100 \\ \hline
Freeway Demand Flow (veh/h) & 4000 & 5500 & 7000 & 6500 & 6000 & 5500 & 4500 \\ \hline
Ramp Demand Flow (veh/h)    & 1000 & 1000 & 1000 & 1000 & 1000 & 1000 & 1000 \\ \hline
\end{tabular}
\end{table}

Table \ref{tab:Proposed framework performance evaluation per Agent} shows MARL-FWC performance evaluation  per agent (the three approaches), where $A$ is defined as the area of the freeway, which starts from the beginning of $A_1$ to the end of $A_3$. And $r_1$, $r_2$, $r_3$ are defined as areas starting from the beginning points of the ramps $1$, $2$, and $3$ to the end points of $A_1$, $A_2$, and $A_3$ respectively. In the IAs, $Agent_1$ tries greedily  to solve its local congestion only regardless of other agents' problems. This leads to overpopulating section $A_2$ of the road, hence increasing the average travel time for $A_2$ and $r_2$ to $112$ and $113$ respectively, and the same for $Agent_3$. In contrary, in coordinated reinforcement learning with Max-plus, the agents optimally cooperate to resolve the congestion problem which leads to 3.2\% advance over the IAs case. Although the CA gives some improvements, this costly solution is not recommended compared to the RL with a Max-plus.
\begin{table}[H]
\caption{MARL-FWC performance evaluation per agent.}
\label{tab:Proposed framework performance evaluation per Agent}
\centering
\begin{tabular}{c|c|c|c|c|c|c|c|c}
\hline
\multicolumn{2}{c}{\multirow{2}{*}{}} & \multicolumn{2}{|c}{$Agent_1$}                    & \multicolumn{2}{|c}{$Agent_2$}                    & \multicolumn{2}{|c|}{$Agent_3$}                  &        \\
\multicolumn{2}{c|}{}                 & $A_1$                      & $r_1$                     & $A_2$                      & $r_2$                     & $A_3$                     & $r_3$                     & $A$      \\ \hline
         & IAs                         & 109.05                  & 89.7                   & 112.06                  & 113.07                 & 97.53                  & 80.92                  & 401.79 \\
         & \multirow{2}{*}{Max-plus}  & \multirow{2}{*}{112.87} & \multirow{2}{*}{69.18} & \multirow{2}{*}{100.34} & \multirow{2}{*}{77.36} & \multirow{2}{*}{94.18} & \multirow{2}{*}{54.93} & 388.85 \\
TT(AVG)(s)  &                            &                         &                        &                         &                        &                        &                        & 3.2\%  \\
         & \multirow{2}{*}{CA}        & \multirow{2}{*}{110.05} & \multirow{2}{*}{70.47} & \multirow{2}{*}{99.09}  & \multirow{2}{*}{77.9}  & \multirow{2}{*}{93.68} & \multirow{2}{*}{54.01} & 385.32 \\
         &                            &                         &                        &                         &                        &                        &                        & 4\%    \\ \hline
\end{tabular}
\end{table}

\begin{figure}[H]
        \centering
        \begin{subfigure}[b]{\linewidth}
				\centering
                \includegraphics[scale=0.4]{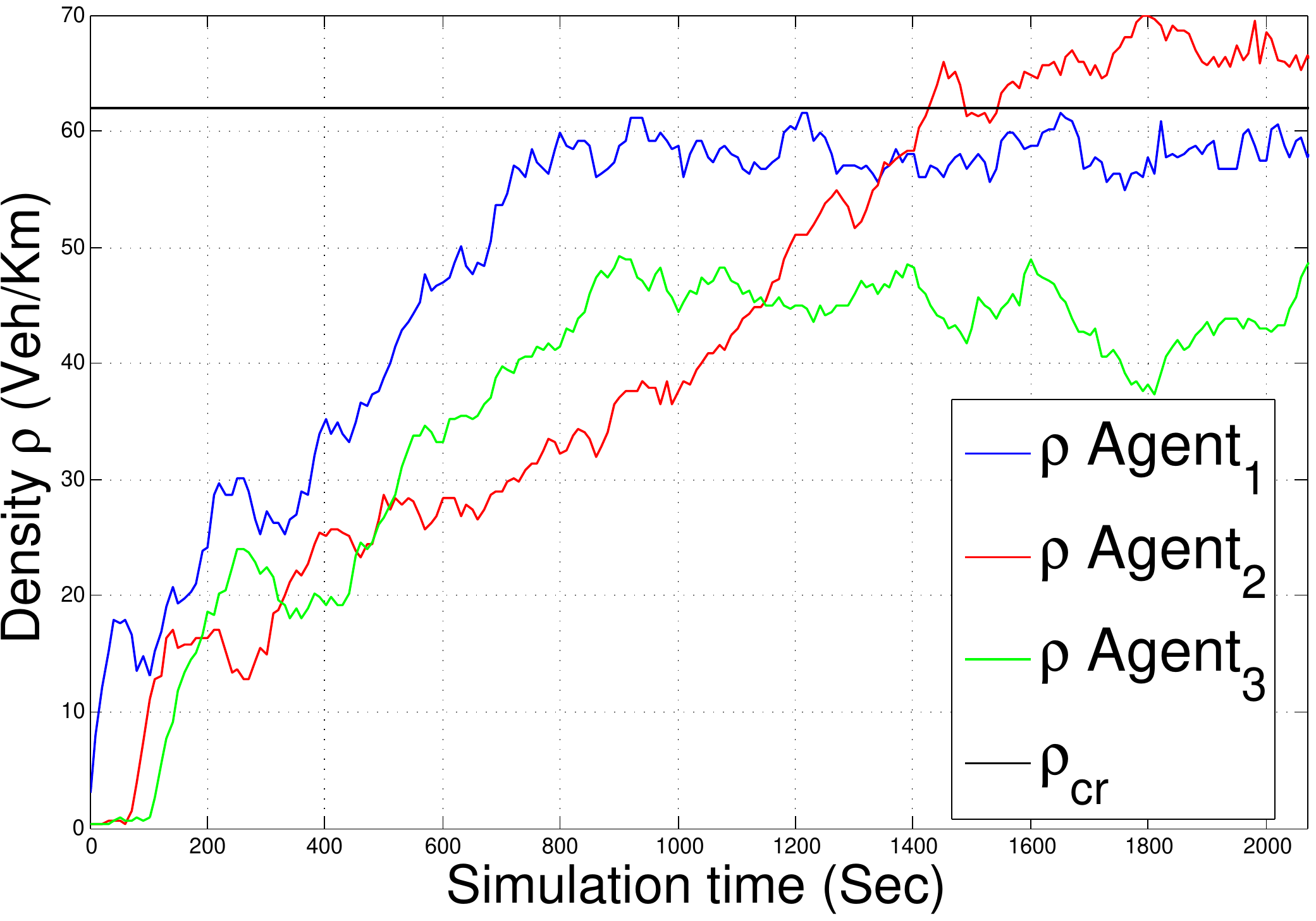}
                \caption{IAs density $\rho$/agent.}
                \label{fig:IAs density per agent}
        \end{subfigure}%
        ~ %add desired spacing between images, e. g. ~, \quad, \qquad etc.
          %(or a blank line to force the subfigure onto a new line)

        \begin{subfigure}[b]{\linewidth}
				\centering
                \includegraphics[scale=0.4]{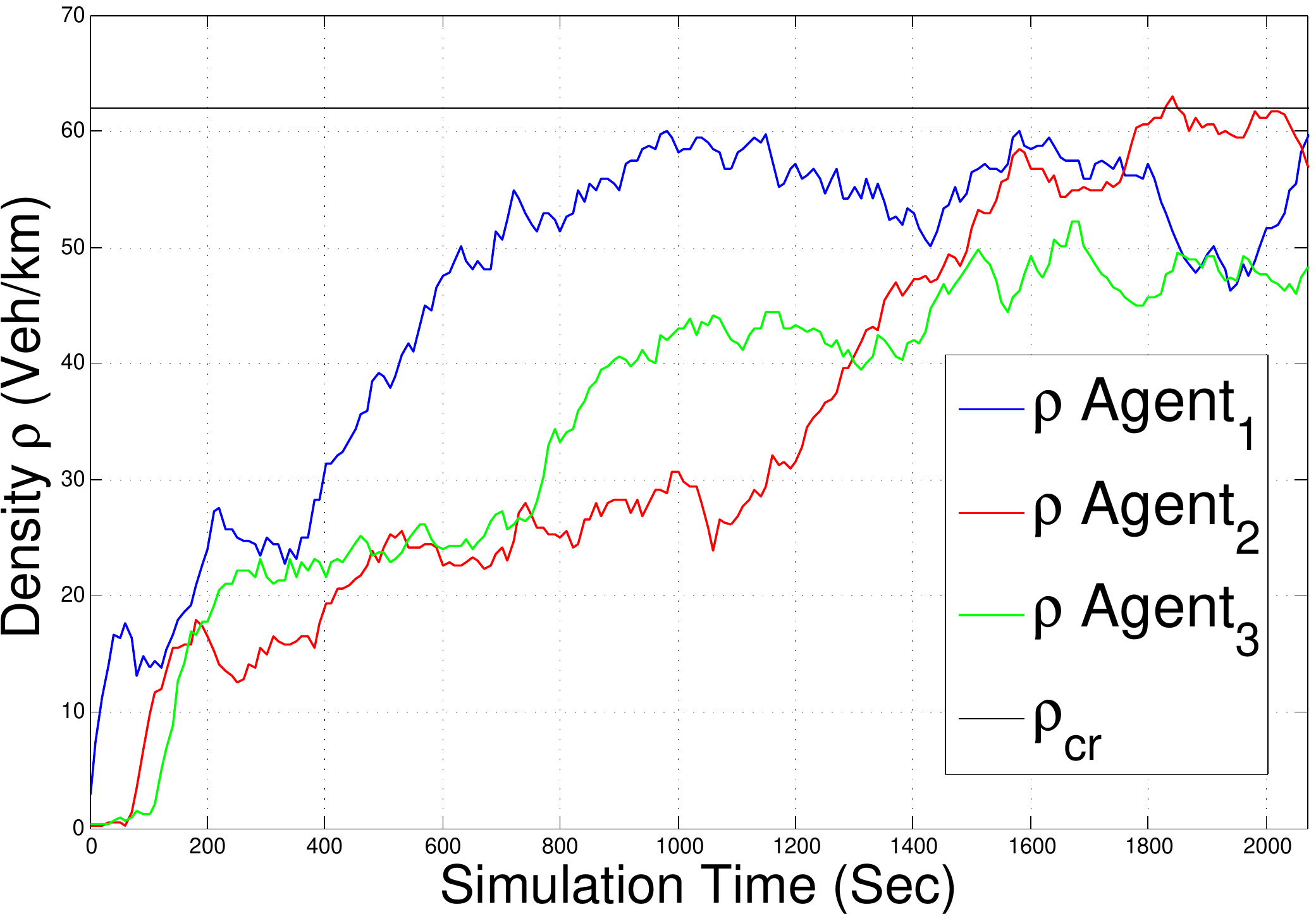}
                \caption{Max-plus density $\rho$/agent.}
                \label{fig:Max-plus density per agent}
        \end{subfigure}

        \begin{subfigure}[b]{\linewidth}
				\centering
                \includegraphics[scale=0.4]{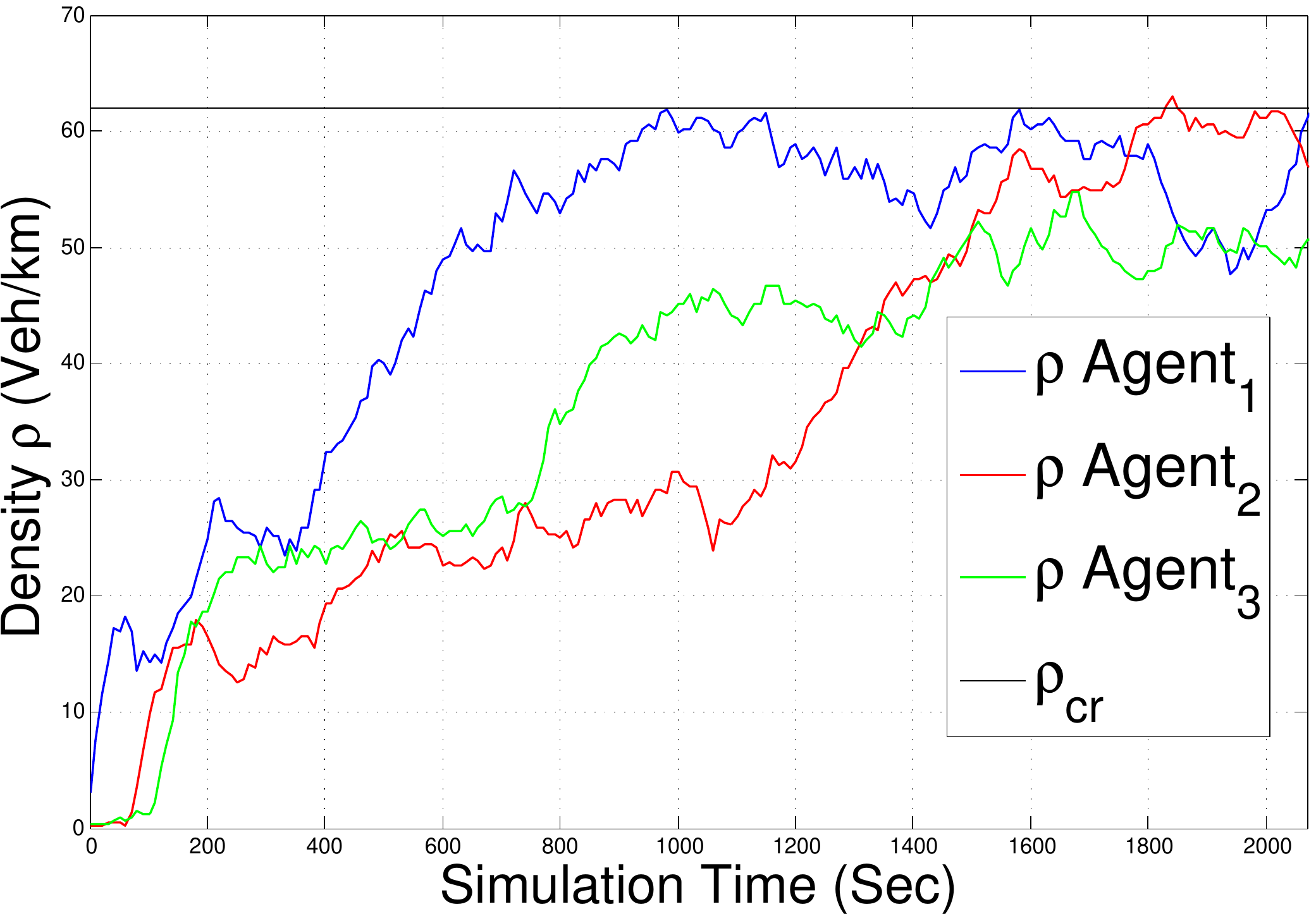}
                \caption{CA density $\rho$/area.}
                \label{fig:CA density per agent}
        \end{subfigure}
\caption{Freeway density for MARL-FWC with the three approaches.}
\label{fig:Freeway density for the proposed framework}
\end{figure}

Line graphs in Fig. \ref{fig:Freeway density for the proposed framework} demonstrate the freeway density associated with each approach of MARL-FWC % (independent agents, coordinated reinforcement learning with Max-plus and centralized agent)
over 2100 seconds of simulation. An important thing here is that, in Fig. \ref{fig:IAs density per agent}, $Agent_1$ tries to maintain its density within a small neighborhood  of the critical density regardless of other agents' performance. This leads to overpopulating section $A_2$ of the road. Hence, there is a dramatic increase in the density of section $A_2$ over the critical ratio between $1400\;s$ and $2100\;s$. This leads to the conclusion that $Agent_2$ does not converge to the optimal solution. Figures \ref{fig:Max-plus density per agent} and \ref{fig:CA density per agent} show that coordinated reinforcement learning with Max-plus and centralized agent successfully keep the density of the freeway at required level over the simulation period.

Table  \ref{tab:Proposed framework performance evaluation}  provides  the  results  obtained  from  applying MARL-FWC  with  all  the  three  approaches and compares it to the base case (no-metering).
It can be seen that  coordinated reinforcement learning with Max-Plus and centralized agent have  shown  considerable improvement over the base case. Centralized agent and cooperative $Q$-learning with Max-plus give $6.5$\% and $6.9$\% in terms of total travel time and $6.74$\% and $7.5$\% in terms of average speed improvements over the base case, respectively.
\begin{table}[h!]
\caption{MARL-FWC performance evaluation.}
\label{tab:Proposed framework performance evaluation}
\centering
\begin{tabular}{c|c|c|c|c}
\hline
                      & No-metering & IAs & Max-plus & Centralized agent \\ \hline
Total travel time (h) & 333         & 331.5   & 311.5    & 308.7         \\
                      &             & 0.5\%   & 6.5\%    & 6.9\%                 \\ \hline
Average speed (km/h)       & 45.25       & 45.49   & 48.5     & 48.65                 \\
                      &             &0.5\%    & 6.74\%    &  7.5\%                \\ \hline
\end{tabular}
\end{table}

An experiment of a dense network with three ramps is provided, however, MARL-FWC can handle whatsoever type of networks with any number of ramps.  In addition, the recommendation for this type of networks and similar ones is to apply the second approach of MARL-FWC which is cooperative $Q$-learning with Max-plus.
%-----------------------------------------------------------------------
\subsection{Adaptive objective function plus DSLs}
\label{subsec:Experimental-results-Adaptive-objective-function-plus-dynamic-speed-limits}
\begin{figure}[H]
\centering
  % Requires \usepackage{graphicx}
  \includegraphics[width=\linewidth]{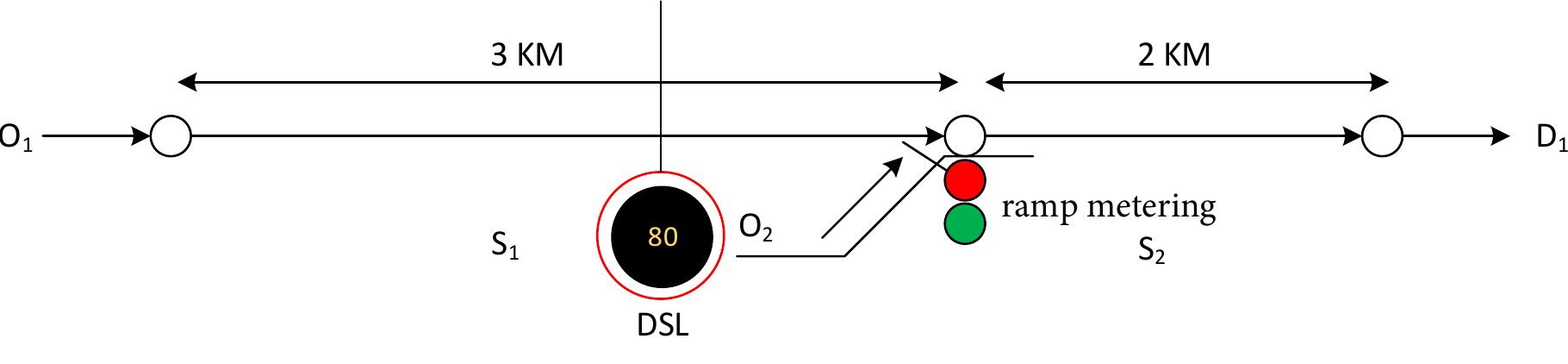}\\
  \caption{Traffic network with DSLs and a metered on-ramp.}
  \label{fig:Traffic network with dynamic speed limits and a metered on ramp}
\end{figure}
The studied network in Fig. \ref{fig:Traffic network with dynamic speed limits and a metered on ramp} consists of a mainstream freeway with two lanes and an on-ramp with one lane. The network consists of two sources of inflow: $O_1$ for the mainstream flow and $O_2$ for the on-ramp flow and it has only one discharge point $D_1$ with unrestricted outflow. The mainstream freeway consists of two sections: $S_1$ and $S_2$. The first section $S_1$ lies before the ramp, it is $4\; km$ long. The second section $S_2$ lies after the ramp, it is $2 \;km$  long. The mainstream freeway capacity is $4000 \;veh/h$ for both lanes, and the on-ramp capacity is $2000 \;veh/h$. The network parameters are taken as follows: $V_{ff}= 60 \;km^{2}/h$, $\rho(0)= 40\;veh/km/lane$, $\rho_{jam}=180\;veh/km/lane$ and $\rho_{cr}=33.5 \;veh/km/lane$.

Table \ref{tab:Demand scenario for the simulation} represents the demand (for both the mainstream freeway and ramp) considered in the simulation experiment over a four hours period. This demand scenario gives the opportunity to study the effect of the on-ramp smart control and DSLs. In the freeway line, there was an increase in the demand flow over the first hour, then the demand flow remains steady at high level near the capacity of the freeway over two hours and half, and finally, the demand flow decreases in one hour to low level. In the ramp line, the demand flow went up over the first hour, then remains at high level near the capacity for half an hour, finally, it dropped to low value and remains steady for two hours and half. The studied network is considered a dense network where the smart control system is highly needed.

\begin{table}[H]
\centering
\caption{Demand scenario for the simulation.}
\label{tab:Demand scenario for the simulation}
\begin{tabular}{l|c|c|c|c|c|c|c|c}
\hline
Simulation Time (h) & 0.5  & 1    & 1.5  & 2    & 2.5  & 3    & 3.5  & 4 \\ \hline
Freeway Flow (veh/h)& 2500 & 3000 & 3500 & 3500 & 3500 & 3500 & 2500 & 1000\\ \hline
Ramp Flow (veh/h)   & 500  & 1000 & 1500 & 500  & 500  & 500  & 500  & 500\\ \hline
\end{tabular}
\end{table}

Table \ref{tab:Proposed adaptive ramp metering plus dynamic speed limits performance evaluation} demonstrates the total time spent (TTS (veh.h.)) between three different strategies. There was a decline in the total time spent, when MARL-FWC for adaptive ramp metering and ramp metering plus DSLs is used as compared to the base case.

\begin{table}[H]
\caption{MARL-FWC for adaptive ramp metering plus DSLs performance evaluation.}
\label{tab:Proposed adaptive ramp metering plus dynamic speed limits performance evaluation}
\centering
\begin{tabular}{c|c|c|c}
\hline
                             & No-metering             & Ramp metering                                                          & Ramp metering plus DSL                                                   \\ \hline
\multirow{2}{*}{TTS(veh.h.)} & \multirow{2}{*}{1240.0} & \multirow{2}{*}{\begin{tabular}[c]{@{}c@{}}1152.0\\ 7\%\end{tabular}} & \multirow{2}{*}{\begin{tabular}[c]{@{}c@{}}1055.6\\ 14.8\%\end{tabular}} \\
                             &                         &                                                                       &                                                                          \\ \hline
\end{tabular}
\end{table}

An important thing to be noticed here is that the new approach for ramp metering plus DSLs have shown considerable improvement over the base case. Adaptive ramp metering and adaptive ramp metering plus DSLs give $7\%$ and $14.8\%$ in terms of total time spent over the base case.
Finally, in on-ramp queue constraint, the DSLs can complement the ramp metering by preventing
traffic break down and maintaining the traffic flow at high level.
%-----------------------------------------------------------------------
%\section{Summary}
%Firstly, this section describes experiments to test the performance of MARL-FWC. It also shows the experimental results. Then, It illustrates the experimental results obtained from testing the performance of MARL-FWC. Finally, it represents the experimental results when testing MARL-FWC modified agent objective function's plus DSLs.

%The experimental results are thoroughly analyzed to study the performance of MARL-FWC using a concrete set of metrics, namely, total travel time, average speed and total time spent while keeping freeway density at optimum level close to the critical ratio for maximum traffic flow.  Our findings prove that MARL-FWC achieved significant enhancement on three features as compared to the base case.

%This section also shows that when the ramp metering could not mitigate the capacity drop at the merging area, DSLs could complement the ramp metering by adjusting the flow rate on the main street via changing the speed limits' values.

\section{Conclusions}
\label{sec:Conclusion}
%\subsection{Conclusions}
%The conclusion goes here.
%The main conclusions of the study may be presented in a short Conclusions section, which may stand alone or form a subsection of a Discussion or Results and Discussion section.
% How the work advances the field from present state of knowledge
%  - Should be clear
%  - Justify your work in the research field
%  - Suggest future experiments

%% The Appendices part is started with the command \appendix;
%% appendix sections are then done as normal sections
%% \appendix

%% \section{}
%% \label{}

In this article, the problem of traffic congestion in freeways at the network level has been addressed. A new system for controlling ramps metering and speed limits has been introduced based on a multi-agent reinforcement-learning framework. MARL-FWC comprises both a MDP modeling technique and an associated cooperative $Q$-learning technique, which is based on payoff propagation (Max-plus algorithm) under the coordination graphs framework.
For MARL-FWC to be as optimal as possible, the approach is evaluated using different modes of operation depending on the network architecture. This framework has been tested on the state-of-the-art VISSIM traffic simulator in a dense practical scenarios.

The experimental results have been thoroughly analyzed to study the performance of MARL-FWC using a concrete set of metrics, namely, total travel time, average speed, and total time spent while keeping freeway density at optimum level close to the critical ratio for maximum traffic flow. The findings have proved that MARL-FWC has achieved significant enhancement on three features as compared to the base case.
It have been also shown that when the ramp metering could not mitigate the capacity drop at the merging area, DSLs could complement the ramp metering by adjusting the flow rate on the main street via changing the speed limits values. The advantages of applying MARL-FWC include saving car fuel, decreasing air pollution, mitigating capacity drop, and saving lives by reducing the chances of car accidents.
%-----------------------------------------------------------------------

%\subsection{Future directions}
%There is currently a joint project with professor Hesham Rakha's lab (http://www.cee.vt.edu/people/rakha.html). This lab has developed a tool called SPD\_CAL for calibrating  steady-state  traffic  stream  and  car-following  models  using  loop  detector  data. In addition, they have contributed a Speed Harmonization (SH) algorithm based on a feedback control system to compute advisory speed limits for vehicles in a SH zone to prevent the breakdown of a downstream bottleneck \cite{lu2015novel}. A field data collected at section of Interstate 66 in Northern Virginia is used for simulation in the INTEGRATION microscopic traffic simulation model to assess the benefits of the SH algorithm. That been said, there is now a field data from Interstate 66, INTEGRATION microscopic traffic simulation, in addition to  calibrated fundamental traffic flow plus SH algorithm for DSLs. The main objective of this joint project is to integrate MARL-FWC as a built-in model in the INTEGRATION microscopic traffic simulation and test its performance against the collected field data of Interstate 66.

%--------------------------------------------------------------------------
%\section{Summary}
%The contributions of the paper are summarized and some possible future research extension are outlined.

\section{Acknowledgements}
%If you'd like to thank anyone, place your comments here
%and remove the percent signs.
%Sincere thanks are due to E-JUST University for guidance and support.
% Collate acknowledgements in a separate section at the end of the article before the references and do not, therefore, include them on the title page, as a footnote to the title or otherwise. List here those individuals who provided help during the research (e.g., providing language help, writing assistance or proof reading the article, etc.).
% Ensures those who helped in the research are recognized
% Include individuals who have assisted your study, including:
% Advisors, Financial supporters, Proofreaders, Typists, Suppliers who may have given materials
This work is mainly supported by the Ministry of Higher Education (MoHE) of Egypt through PhD fellowship awarded to Dr. Ahmed Fares.
This work is supported in part by the Science and Technology Development Fund (STDF), Project ID 12602 ''Integrated Traffic Management and Control System'', and by E-JUST Research Fellowship awarded to Dr. Mohamed A. Khamis.
%\end{acknowledgements}
%% References
%%

\bibliographystyle{vancouver-authoryear}
\bibliography{references}

\begin{thebibliography}{28}
\providecommand{\natexlab}[1]{#1}
\providecommand{\url}[1]{\texttt{#1}}
\providecommand{\urlprefix}{}

\bibitem[{Bertsekas(1996)Bertsekas, Dimitri P}]{bertsekas1995dynamic}
Bertsekas DP.
\newblock Dynamic programming and optimal control, vol.~1.
\newblock Athena Scientific Belmont, Massachusetts; 1996.

\bibitem[{Ernst et~al.(2009)Ernst, Damien and Glavic, Mevludin and Capitanescu,
  Florin and Wehenkel, Louis}]{ernst2009reinforcement}
Ernst D, Glavic M, Capitanescu F, Wehenkel L.
\newblock Reinforcement learning versus model predictive control: a comparison
  on a power system problem.
\newblock Systems, Man, and Cybernetics, Part B: Cybernetics, IEEE Transactions
  on 2009;39(2):517--529.

\bibitem[{Fares and Gomaa(2014)Fares, A. and Gomaa, W.}]{6871097}
Fares A, Gomaa W.
\newblock Freeway ramp-metering control based on Reinforcement learning.
\newblock In: Control Automation (ICCA), 11th IEEE International Conference on;
  2014. p. 1226--1231.

\bibitem[{Fares and Gomaa(2015)Fares, Ahmed and Gomaa, Walid}]{123456}
Fares A, Gomaa W.
\newblock Multi-Agent Reinforcement Learning Control for Ramp Metering.
\newblock In: Selvaraj H, Zydek D, Chmaj G, editors. Progress in Systems
  Engineering, vol. 330 of Advances in Intelligent Systems and Computing
  Springer International Publishing; 2015.p. 167--173.

\bibitem[{Feng et~al.(2011)Chen Feng and Jia Yuanhua and Li Jian and Yi Huixin
  and Niu Zhonghai}]{5720945}
Feng C, Yuanhua J, Jian L, Huixin Y, Zhonghai N.
\newblock Design of Fuzzy Neural Network Control Method for Ramp Metering.
\newblock In: Measuring Technology and Mechatronics Automation (ICMTMA), 2011
  Third International Conference on, vol.~1; 2011. p. 966--969.

\bibitem[{Ghods et~al.(2009)Ghods, A.H. and Kian, A. and Tabibi, M.}]{5117656}
Ghods AH, Kian A, Tabibi M.
\newblock Adaptive freeway ramp metering and variable speed limit control: a
  genetic-fuzzy approach.
\newblock Intelligent Transportation Systems Magazine, IEEE 2009;1(1):27--36.

\bibitem[{Guestrin et~al.(2001)Guestrin, Carlos and Koller, Daphne and Parr,
  Ronald}]{guestrin2001multiagent}
Guestrin C, Koller D, Parr R.
\newblock Multiagent planning with factored MDPs.
\newblock In: NIPS, vol.~1; 2001. p. 1523--1530.

\bibitem[{Guestrin(2003)Guestrin, Carlos Ernesto}]{guestrin2003planning}
Guestrin CE.
\newblock Planning under uncertainty in complex structured environments.
\newblock PhD thesis, Stanford University; 2003.

\bibitem[{Kaelbling and Moore(1996)Kaelbling, Leslie Pack, Michael L. Littman,
  and Andrew W. Moore}]{kaelbling1996reinforcement}
Kaelbling MLL Leslie~Pack, Moore AW.
\newblock Reinforcement learning: A survey.
\newblock Journal of artificial intelligence research 1996;p. 237--285.

\bibitem[{Khamis and Gomaa(2012)Mohamed A. Khamis and Walid
  Gomaa}]{Khamis2012EnhancedMultiObjectiveTrafficControl}
Khamis MA, Gomaa W.
\newblock Enhanced Multiagent Multi-Objective Reinforcement Learning for Urban
  Traffic Light Control.
\newblock In: Proc. {IEEE} 11th International Conference on Machine Learning
  and Applications ({ICMLA} 2012) Boca Raton, Florida; 2012. p. 586--591.

\bibitem[{Khamis and Gomaa(2014)Mohamed A. Khamis and Walid
  Gomaa}]{Khamis2014AdaptiveMultiObjectiveReinforcementLearning}
Khamis MA, Gomaa W.
\newblock {Adaptive multi-objective reinforcement learning with hybrid
  exploration for traffic signal control based on cooperative multi-agent
  framework}.
\newblock {Journal of Engineering Applications of Artificial Intelligence} 2014
  March;29:134--151.

\bibitem[{Khamis and Gomaa(2015)Mohamed A. Khamis and Walid
  Gomaa}]{Khamis2015ComparativeAssessmentMLScoringFunctionsOnPDBbind2013}
Khamis MA, Gomaa W.
\newblock {Comparative Assessment of Machine-Learning Scoring Functions on
  PDBbind 2013}.
\newblock {Engineering Applications of Artificial Intelligence} 2015;p.
  136--151.

\bibitem[{Khamis et~al.(2015)Mohamed A. Khamis and Walid Gomaa and Walaa F.
  Ahmed}]{Khamis2015MachineLearningInComputationalDocking}
Khamis MA, Gomaa W, Ahmed WF.
\newblock {Machine learning in computational docking}.
\newblock {Artificial Intelligence In Medicine} 2015;63:135--152.

\bibitem[{Khamis et~al.(2012{\natexlab{a}})Mohamed A. Khamis and Walid Gomaa
  and Ahmed El-Mahdy and Amin Shoukry}]{Khamis2012AdaptiveTrafficControl}
Khamis MA, Gomaa W, El-Mahdy A, Shoukry A.
\newblock Adaptive Traffic Control System Based on Bayesian Probability
  Interpretation.
\newblock In: Proc. {IEEE} 2012 Japan-Egypt Conference on Electronics,
  Communications and Computers ({JEC-ECC} 2012) Alexandria, Egypt; 2012. p.
  151--156.

\bibitem[{Khamis et~al.(2012{\natexlab{b}})Mohamed A. Khamis and Walid Gomaa
  and Hisham El-Shishiny}]{Khamis2012MultiObjectiveTrafficControl}
Khamis MA, Gomaa W, El-Shishiny H.
\newblock Multi-Objective Traffic Light Control System based on Bayesian
  Probability Interpretation.
\newblock In: Proc. {IEEE} 15th International Conference on Intelligent
  Transportation Systems ({ITSC} 2012) Anchorage, AK; 2012. p. 995--1000.

\bibitem[{Kok and Vlassis(2006)Kok, Jelle R and Vlassis,
  Nikos}]{kok2006collaborative}
Kok JR, Vlassis N.
\newblock Collaborative multiagent reinforcement learning by payoff
  propagation.
\newblock The Journal of Machine Learning Research 2006;7:1789--1828.

\bibitem[{Kschischang et~al.(2001)Kschischang, Frank R and Frey, Brendan J and
  Loeliger, H-A}]{kschischang2001factor}
Kschischang FR, Frey BJ, Loeliger HA.
\newblock Factor graphs and the sum-product algorithm.
\newblock Information Theory, IEEE Transactions on 2001;47(2):498--519.

\bibitem[{Li and Liang(2009)Jianye Li and Xinrong Liang}]{5358206}
Li J, Liang X.
\newblock Freeway ramp control based on single neuron.
\newblock In: Intelligent Computing and Intelligent Systems, 2009. ICIS 2009.
  IEEE International Conference on, vol.~2; 2009. p. 122--125.

\bibitem[{Liang et~al.(2010)Xinrong Liang and Jianye Li and Nongzhen
  Luo}]{5678377}
Liang X, Li J, Luo N.
\newblock Single Neuron Based Freeway Traffic Density Control via Ramp
  Metering.
\newblock In: Information Engineering and Computer Science (ICIECS), 2010 2nd
  International Conference on; 2010. p. 1--4.

\bibitem[{Papageorgiou et~al.(1991)Papageorgiou, Markos and Hadj-Salem, Habib
  and Blosseville, Jean-Marc}]{papageorgiou1991alinea}
Papageorgiou M, Hadj-Salem H, Blosseville JM.
\newblock ALINEA: A local feedback control law for on-ramp metering.
\newblock Transportation Research Record 1991;(1320).

\bibitem[{Pearl(1988)Pearl, Judea}]{pearl1988probabilistic}
Pearl J.
\newblock Probabilistic reasoning in intelligent systems: networks of plausible
  inference.
\newblock Morgan Kaufmann; 1988.

\bibitem[{Prendinger et~al.(2013)Prendinger, H. and Gajananan, K. and Bayoumy
  Zaki, A. and Fares, A. and Molenaar, R. and Urbano, D. and van Lint, H. and
  Gomaa, W.}]{6582403}
Prendinger H, Gajananan K, Bayoumy~Zaki A, Fares A, Molenaar R, Urbano D,
  et~al.
\newblock Tokyo Virtual Living Lab: Designing Smart Cities Based on the {3D}
  Internet.
\newblock Internet Computing, IEEE 2013 Nov;17(6):30--38.

\bibitem[{Vlassis et~al.(2004)Vlassis, Nikos and Elhorst, Reinoud and Kok,
  Jelle R}]{vlassis2004anytime}
Vlassis N, Elhorst R, Kok JR.
\newblock Anytime algorithms for multiagent decision making using coordination
  graphs.
\newblock In: Systems, Man and Cybernetics, 2004 IEEE International Conference
  on, vol.~1 IEEE; 2004. p. 953--957.

\bibitem[{Wainwright et~al.(2004)Wainwright, Martin and Jaakkola, Tommi and
  Willsky, Alan}]{wainwright2004tree}
Wainwright M, Jaakkola T, Willsky A.
\newblock Tree consistency and bounds on the performance of the max-product
  algorithm and its generalizations.
\newblock Statistics and Computing 2004;14(2):143--166.

\bibitem[{Watkins and Dayan(1992)Watkins, Christopher JCH and Dayan,
  Peter}]{watkins1992q}
Watkins CJ, Dayan P.
\newblock Q-learning.
\newblock Machine learning 1992;8(3-4):279--292.

\bibitem[{Yedidia et~al.(2003)Yedidia, Jonathan S and Freeman, William T and
  Weiss, Yair}]{yedidia2003understanding}
Yedidia JS, Freeman WT, Weiss Y.
\newblock Understanding belief propagation and its generalizations.
\newblock Exploring artificial intelligence in the new millennium
  2003;8:236--239.

\bibitem[{Yu et~al.(2012)Yu, X.F. and Xu, W.L. and Alam, F. and Potgieter, J.
  and Fang, C.F.}]{6484604}
Yu XF, Xu WL, Alam F, Potgieter J, Fang CF.
\newblock Genetic fuzzy logic approach to local ramp metering control using
  microscopic traffic simulation.
\newblock In: Mechatronics and Machine Vision in Practice (M2VIP), 2012 19th
  International Conference; 2012. p. 290--297.

\bibitem[{Zaky et~al.(2015)Ahmed Bayoumy Zaky and Walid Gomaa and Mohamed A.
  Khamis}]{Zaky2015CarFollowingMarkovRegimeClassification}
Zaky AB, Gomaa W, Khamis MA.
\newblock Car Following Markov Regime Classification and Calibration.
\newblock In: Proceedings of the IEEE 14th International Conference on Machine
  Learning and Applications (ICMLA 2015) Miami, Florida, USA: IEEE; 2015. .

\end{thebibliography}

\end{document}